\documentclass[letterpaper,twocolumn,10pt]{article}
\usepackage{usenix,epsfig,endnotes}
\usepackage{times}
\usepackage{dingbat}
\usepackage{makecell}
\usepackage{flushend}

\usepackage{pifont}
\newcommand{\xmark}{\ding{55}}%

\usepackage{amsmath}
\usepackage{hyperref}
\usepackage{multirow}
\usepackage{booktabs}
\usepackage{grffile}
\usepackage{todonotes}
\usepackage{mathtools}
\usepackage[linesnumbered,ruled,vlined]{algorithm2e}
\usepackage{subcaption}
\usepackage{xcolor}
\usepackage{hyperref}
\usepackage{url} 
\usepackage[small,compact]{titlesec}
\usepackage[font=footnotesize,format=plain,labelfont=bf,textfont=it]{caption}
\usepackage{xspace} 
\usepackage{balance} 
\usepackage{listings} 
\SetAlFnt{\small\sffamily}

\DeclarePairedDelimiter\floor{\lfloor}{\rfloor}

\newenvironment{itemtight}{
 \begin{list}{$\bullet$\hfill}
  {\setlength{\parsep}{0ex}\setlength{\itemsep}{0ex}
   \setlength{\labelwidth}{0.10in}\setlength{\labelsep}{0.05in}
   \setlength{\leftmargin}{0.15in}\setlength{\rightmargin}{0.0in}
  }
}{
 \end{list}
}

\begin{document}

\date{}

\title{\Large \bf \textsc{Mithril}:  Mining Sporadic Associations for Cache Prefetching}

\author{
{\rm Juncheng Yang$^*$},
{\rm Reza Karimi$^*$},
{\rm Trausti S\ae mundsson$^\ddagger$},
{\rm Avani Wildani$^*$},
{\rm Ymir Vigfusson$^{*,\dagger}$}\\
$^*$Emory University, $^\dagger$Reykjavik University, $^\ddagger$CachePhysics Inc.
} 

\maketitle


\newcommand{\alg}{\textsc{Mithril}}
\newcommand{\Alg}{\textsc{Mithril}}
\newcommand{\mAmp}{\textsc{Mithril-AMP}}
\newcommand{\mLRU}{\textsc{Mithril-LRU}}
\newcommand{\amp}{\textsc{Amp}}
\newcommand{\Amp}{\textsc{Amp}}
\newcommand{\LRU}{\textsc{LRU}}
\newcommand{\FIFO}{\textsc{FIFO}}
\newcommand{\PG}{\textsc{Probability Graph}}
\newcommand{\PGs}{\textsc{PG}}

\newcommand{\placeholder}{not\xspace} 

\newcommand{\lookaheadWindow}{\textit{lookahead range }$\Delta$}
\newcommand{\lookaheadWindowS}{\textit{lookahead range }$\Delta$}

\newcommand{\minSup}{\textit{minimum support }$R$}
\newcommand{\maxSup}{\textit{maximum support }$S$}
\newcommand{\prefetchingListSize}{\textit{prefetching list size }$P$}
\newcommand{\maxMetaDataSize}{\textit{maximum metadata size }$M$} 

\newcommand{\LookaheadWindow}{\textit{Lookahead range }$\Delta$}
\newcommand{\MinSup}{\textit{Minimum support }$R$}
\newcommand{\MaxSup}{\textit{Maximum support }$S$}
\newcommand{\PrefetchingListSize}{\textit{Prefetching list size }$P$}
\newcommand{\MaxMetaDataSize}{\textit{Maximum metadata size }$M$}

\newcommand{\mt}{\textit{mining table}}
\newcommand{\rt}{\textit{recording table}}
\newcommand{\pt}{\textit{prefetching table}}
\newcommand{\Mt}{\textit{Mining table}}
\newcommand{\Rt}{\textit{Recording table}}
\newcommand{\Pt}{\textit{Prefetching table}}

\newenvironment{myitemize}{
\begin{itemize}
  \setlength{\itemsep}{0pt}
  \setlength{\parskip}{-0pt}
  \setlength{\parsep}{-0pt}
}{\end{itemize}}

\vskip-0.4in

\subsection*{Abstract}
The growing pressure on cloud application scalability has
accentuated storage performance as a critical bottleneck.
Although cache replacement algorithms have been extensively studied, 
cache prefetching -- reducing latency by retrieving items before they are actually requested -- remains an underexplored area. Existing approaches to history-based prefetching, in particular, provide too few benefits for real systems for the resources they cost. 

We propose \alg{}, a prefetching layer that efficiently exploits historical patterns in cache request associations. \alg{} is inspired by sporadic association rule mining and only relies on the timestamps of requests. Through evaluation of 135 block-storage traces, we show that \alg{} is effective, giving an average of a 55\% hit ratio increase over \LRU{} and \PG{}, a 36\% hit ratio gain over \amp{} at reasonable cost. We further show that \alg{} can supplement any cache replacement algorithm and be readily integrated into existing systems. 
Furthermore, we demonstrate the improvement comes from \alg{} being able to capture mid-frequency blocks.

\section{Introduction}
\label{sec: intro}

As cloud tenants use increasing volumes of data, the pressure mounts on the underlying storage systems to prevent
high access latencies for end-users. 
The prevalent techniques for mitigating block storage access latencies are 
to cache recently accessed blocks \cite{Waldspurger2015Shards}, 
and to prefetch blocks into the cache in advance of anticipated accesses 
\cite{yang2016tombolo,jiang2013prefetching}. 

Current approaches to cache prefetching can be divided into two schools. 
On one hand, sequential prefetching techniques (such as AMP \cite{Gill2007AMPFull}) anticipate access to consecutive block identifiers,
but rely on block I/O with progressive data layout. 
On the other hand, history-based prefetching seeks to find and exploit deep correlations among past accesses but normally at substantial computational cost \cite{li2004cminer}. 
To mitigate overhead and to make caching and prefetching more effective, several applications choose to provide additional hints \cite{soundararajan2008quickmine} with each access \cite{chang99,gniady04,wong02,li2004cminer,li2005mining}. 
Passing extra information, however, requires restructuring, reorganization or modification to the software stack \cite{soundararajan2008quickmine}, 
and is infeasible in scenarios where parts of the stack is proprietary. 

We argue that to avoid becoming a latency bottleneck, modern block storage systems need general prefetching techniques that fulfill the following criteria.
\begin{myitemize}
\item
\textbf{Exploit history.} Various lower layers of storage systems perform sequential prefetching so the focus should be on the more spatially and temporally sophisticated patterns of reuse.
\item
\textbf{Low overhead.}
The methods must be simple, on-line and impose low time and space overhead. 
\item
\textbf{Backward compatible.}
The methods should implement standard legacy interfaces and treat other parts of the storage system as a black-box. 
\end{myitemize}
Existing approaches fall short of one or more of these goals: probability graphs and variants incur intensive space or computation overhead  \cite{griffioen1994probabilityGraph,li2004cminer,yang2016tombolo}; QuickMine is an online algorithm but relies on hints from the applications
through modified interfaces \cite{soundararajan2008quickmine} with extra hints from system or applications.


In this paper, we propose \alg{}, a lightweight online history-based prefetching layer which meets all of the goals. 
\Alg{} can be coupled with any existing caching layer, even composed with a sequential prefetching layer such as AMP \cite{Gill2007AMPFull}. 
\Alg{} 
harnesses several concepts from sporadic association rule mining \cite{Koh2005Sporadic} from the data mining literature. 
The central idea behind \alg{} is to track temporal associations between only those blocks whose access patterns are moderately frequent. 
Intuitively, items that are accessed regularly are already handled by an underlying caching system, such as \LRU{}, whereas items that are rarely accessed need not occupy the precious cache memory.
\Alg{} detects associated access patterns between pairs of blocks without relying on application-level hints. 
In contrast to other history-based prefetching algorithms \cite{li2004cminer,li2005mining,griffioen1994probabilityGraph}, \alg{} is
able to discover relationships between interleaved requests that are not consecutive -- a ubiquitous scenario in modern multi-tenant storage systems -- without incurring high computation overhead. 
The focus of this paper is on exploiting patterns in block I/O workloads, but evidence shows that \alg{} works on proxy workloads as well. 
We evaluated \alg{} through experiments on traces from a commercial I/O caching analytics service, CloudPhysics \cite{Waldspurger2015Shards}, 
as well as file system traces from Microsoft Research (MSR) \cite{Narayanan2008MSR}. 
We found that \alg{} boosts the cache hit ratio by up to 7$\times$ over typical cache strategies (\LRU{}) improves over the state-of-the-art sequential prefetching algorithm \amp{} by 36\% on average. 

Our paper makes three contributions.
\begin{itemtight} 
\item A design of a history-based prefetching layer \alg{} that leverages a novel, low-overhead algorithm to mine for regularity in request timestamps in an optimized manner. 
\item A trace-driven experimental evaluation of \alg{} on 135 traces from real storage systems, showing that our \alg{} layer effectively discovers block associations for prefetching. On average, \alg{} increased hit ratio by 56\% in over \LRU{}, and 36\% over \Amp{}. We also measured the latency of \alg{} on a real system. 
\item A demonstration that \alg{} discovers associations between separated blocks from interleaved applications, and the power of \alg{} comes from capturing mid-frequency blocks. 
\end{itemtight}

\section{Background and Motivation}
\label{sec: background}




\begin{table*}[t!]
\caption{Comparison of common prefetching approaches. Overhead and improvement is measured over over \LRU{} on 135 traces (see Sec.~\ref{sec:eval}). \emph{Backward compatible} algorithms require no hints or changes to legacy interfaces. \emph{General} approaches generalize beyond block I/O traces. 
}
\begin{tabular}{l|ccccccc}
\textbf{Algorithm}
 &  \makecell{\textbf{Time} \\ \textbf{overhead}}
 & \makecell{\textbf{Space} \\ \textbf{overhead}}
 & \makecell{\textbf{Avg.~hit ratio} \\ \textbf{improvement}}
& \makecell{\textbf{Max.~hit ratio} \\ \textbf{improvement}}
 & \textbf{Online}
& \makecell{\textbf{Backward} \\ \textbf{compatible}} 
 & \textbf{General}
 \\
 \hline
\Amp{} \cite{gill2007amp}  & Low & Low & 12.2\% & 139\% & \checkmark & \checkmark  & \xmark \\ 
\PGs{} \cite{griffioen1994probabilityGraph}  & Low & High & 4.1\% & 156\% & \checkmark & \checkmark & \checkmark \\
C-Miner \cite{li2004cminer} &High & Moderate & \emph{N/A} & \emph{N/A} & \xmark & \checkmark & \checkmark \\ 
QuickMine \cite{soundararajan2008quickmine} &  Moderate & Moderate & \emph{N/A} & \emph{N/A} & \checkmark & \xmark & \checkmark \\
\Alg{} &  Moderate & Moderate & 54.3\% &740\% & \checkmark & \checkmark & \checkmark \\
\end{tabular}
\label{tb:comparison}
\end{table*}

Caching has been widely studied over the past 70 years. The standard algorithm of evicting the least-recently-used elements (LRU) has seen some structural improvements over the years \cite{jiang2002LIRS,zhou2001MultiQueue,Megiddo2003ARC, Tang2015RIPQ}. 
A complementary approach is to prefetch data into the cache before it is used, typically either based on sequential or history-based patterns \cite{soundararajan2008quickmine,yang2016tombolo}.
We argue there is room for improvement for prefetching on block I/O workloads.

\noindent
\textbf{Sequential prefetching is exploited at lower layers.} 

In sequential prefetching, the storage server exploits spatial locality in the I/O request stream by retrieving a batch of consecutive blocks upon detecting a sequential access pattern \cite{gill2007amp,li2008tap}.
Static size sequential prefetching is well-understood, simple to implement and has seen long deployment, but can cause cache pollution in workloads where the  sequential correlation length is variable and affect accuracy. 

Cloud environments, however, exhibit high levels of concurrency. This  results in I/O workloads where  multiple applications interleave I/O accesses that break the continuity of consecutive access patterns \cite{soundararajan2008quickmine}.
Adaptive algorithms such as \amp{} (Adaptive Multi-stream Prefetching) \cite{gill2007amp, Gill2007AMPFull} and \textsc{TaP} (Table-based Prefetching) \cite{li2008tap} dynamically decide when and how much to prefetch. 
\amp{}, for instance, dynamically adjusts the number of pages to be prefetched to prevent both cache pollution and prefetch wastage when the  requests stream is interleaved. \amp{} increases its prefetch degree if the prefetched blocks are waited on by system, and decreased if prefetched blocks are evicted. 
Unlike other prefetching algorithms, which use read cache to detect sequential streams, \textsc{TaP} uses a table to detect sequentiality and track longer history. Thus, \textsc{TaP} outperforms \amp{} on interleaved workloads and at small cache sizes. 

Sequential prefetching has been widely deployed and commonly used in operating systems \cite{McKusick1984FFS, Baker1991MDF}, databases \cite{Teng1984MID} and storage controllers \cite{gill2005sarc}. The ubiquity and success of the approach at lower layers, however,  makes the approach less attractive for higher layers in the storage hierarchy, such as at the virtualization layer. 
The length of contiguous I/O sequences, furthermore, tend to be short at the lowest levels of the storage hierarchy \cite{yang2016tombolo} due to virtualization, multi-tenancy, disk encryption and sophisticated file system layouts. 
Together, these trends reduce the effectiveness of sequential prefetching on today's storage system workloads. 

\textbf{History-based prefetching has been expensive.}
History-based prefetching, in contrast, tolerates discontinuity across repeating patterns at the cost of added complexity and overhead \cite{jiang2013prefetching,griffioen1994probabilityGraph}.
One approach is to generate a directed probability graph over data items, where  arc denotes one item is likely
accessed before the other, and each arc is weighed by the probability of such an access \cite{griffioen1994probabilityGraph,amer2002group,gu2006nexus,yang2016tombolo}. 
Many systems following this direction prevent graph metadata from becoming unwieldy by operating at the file-level instead of the block-level \cite{griffioen1994probabilityGraph,amer2002group,gu2006nexus}, which has inherent limitations \cite{jiang2013prefetching}.

Another take on history-based prefetching is to leverage data mining techniques to identify repeating sequences.
By mapping a block to an item, using frequent sequence mining on the request sequence, we can obtain frequent subsequences in an access stream. A frequent subsequence implies that the involved blocks are frequently accessed together. In other words, frequent subsequences are good indicators for block correlations in a storage system. C-Miner \cite{li2004cminer} and QuickMine \cite{soundararajan2008quickmine} employ this technique to discover block correlations in storage systems. However, precise data mining technique comes with high overhead. C-Miner only runs offline due to its overhead. QuickMine improves on the issue by tagging each application I/O block request with a context identifier corresponding to the higher level application context (\emph{e.g.}, a web interaction, database transaction, \emph{etc.}). 
The tag enables the request sequence to be split before mining, thus making computation overhead manageable. The key novelty of QuickMine lies in detecting and leveraging block correlations within logical application contexts. Nevertheless, it depends on explicit contextual hints from applications, which makes it hard to deploy and impractical for legacy systems.

Current history-based prefetching approaches may capture complex access patterns, but require either explicit contextual information from applications or suffer from high runtime overheads.

\textbf{Temporal block associations should be exploited.} 
Block associations are common in storage systems \cite{li2004cminer}. Sequential prefetching aims to exploit spatially associated blocks, yet temporal associations are equally important for prefetching. Lacking a fast history-based approach, our goal in this paper is thus to \emph{efficiently find temporally associated blocks}. Table \ref{tb:comparison} shows the main algorithms for comparison.


\section{Data Mining Techniques}
\label{sec:sporadic}
In search for an approach to efficiently gather history for cache requests to improve on prefetching, we  survey relevant problems from the data mining literature before
describing our approach.

\subsection{Sporadic Association Rule Mining}
Frequent itemset mining aims to discover which items co-occur frequently in a transaction database. In this field, a group of items is called an \textit{itemset}, and the number of transactions containing this itemset in the database is called \textit{support}. Suppose we have a transaction database. We say an itemset $A$ is \textit{frequent} if its support \textit{$support_A$} is larger than or equal to some threshold, \minSup{}.

Association rule mining is the discovery of a relationship between items $a$ and $b$ in a frequent itemset discovered from the previous step. We say $a\Rightarrow b$ if the probability of $b$ appearing given $a$ is above a threshold. 



Sporadic association rule mining focuses on associations composed of mid-frequency items. It usually consists of three steps. In the first step, frequent itemsets are generated like before. The following step filters out highly frequent itemsets, which are defined as those appearing more than \maxSup{} times; and the frequent itemsets left are called sporadic frequent itemsets. In the third step, association rule mining is used to generate association rules from the sporadic frequent itemsets. By definition, only mid-frequency itemsets and association rules are discovered during the process \cite{han2011dataMiningBook}. 

\subsection{Generalizing to Block Associations}
Let $B=\{b_{1}, b_{2}, \dots, b_{n}\}$ be a sequence of cache block I/O requests. In order to conduct effective prefetching, we need to identify pairs of requests $\{b_{x}, b_{y}\}$ that are likely to co-occur but not too frequently to be captured by the underlying cache. Notice the similarity to sporadic association rule mining: both try to find related items that appear close by and have mid-range frequency. 

To discover such an association, the basic idea is to apply an existing available sporadic association rule mining algorithm \cite{Koh2005Sporadic}. 
However, there are several challenges. A typical storage system can serve up to  billions of requests per day, resulting in an unmanageably long request $B$. In order to conduct sporadic association rule mining on the data, we need to transform the request sequence into a transaction database as the first step. 

The first difficulty is determining how to split $B$ into transactions. One approach is to split $B$ according to wall clock time, for example, splitting requests into transactions every five seconds. Another approach is to split $B$ using some fixed number of requests per transaction, \emph{e.g.}, group every 20 requests into a transaction. However, both approaches result in information loss, because no evidence indicates that two requests separated in different transactions are not associated. Recall that only items in the same transaction can be discovered as frequent itemsets and as being potentially associated. To address this problem, Soundararajan's approach \cite{soundararajan2008quickmine} using a context given by an application to split the sequence is effective but requires changes to the underlying system to obtain such hints, which sacrifices the generality for which \alg{} is designed. 

The second difficulty comes from the high time and space complexity of the currently available sporadic association rule mining algorithms. Koh \cite{Koh2005Sporadic} proposed an optimization for mining sporadic association rules using \textsc{Apriori-Inverse}. Their algorithm, however, still requires two phases: mining all sporadically frequent itemsets and discovering sporadic association rules. Although the algorithm avoids generating and storing highly frequent itemsets, \textsc{Apriori-Inverse} still needs to store and count all possible associated pairs at significant computation and storage overheads, as confirmed using the SPMF library\cite{spmf2014}. 


To efficiently discover associations between requests without requiring extra application-level hints, we propose the \alg{} prefetching layer, whose algorithm provides a fast approximation to sporadic association rule mining. 

\section{Design of \Alg{}}
\label{sec: design}

\begin{figure}[!t]
\begin{center}
\includegraphics[width=.9\linewidth]{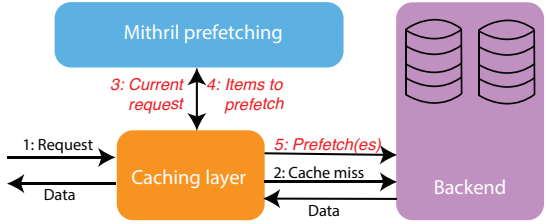}
\caption{Schematic of the \Alg{} prefetching layer.}
\label{fig: scheme1}
\end{center}
\end{figure}

\alg{} is a prefetching layer between the existing caching layer and the backend, as shown in Figure~\ref{fig: scheme1}. Without \alg{}, when a request arrives, it first touches the caching layer; if it is a cache hit, it returns directly from the cache, otherwise, as a cache miss, the application or caching layer needs to go to the backend to fetch the item. When \alg{} is added, 
when a request arrives, \alg{} records the request for mining, checks the potential prefetching list, and sends the request(s) to the caching layer for prefetching. 

\subsection{\Alg{} Mining} 
\label{sec: desAlgorithm}
\begin{figure*}[!ht]
\begin{center}
\includegraphics[width=0.8\linewidth]{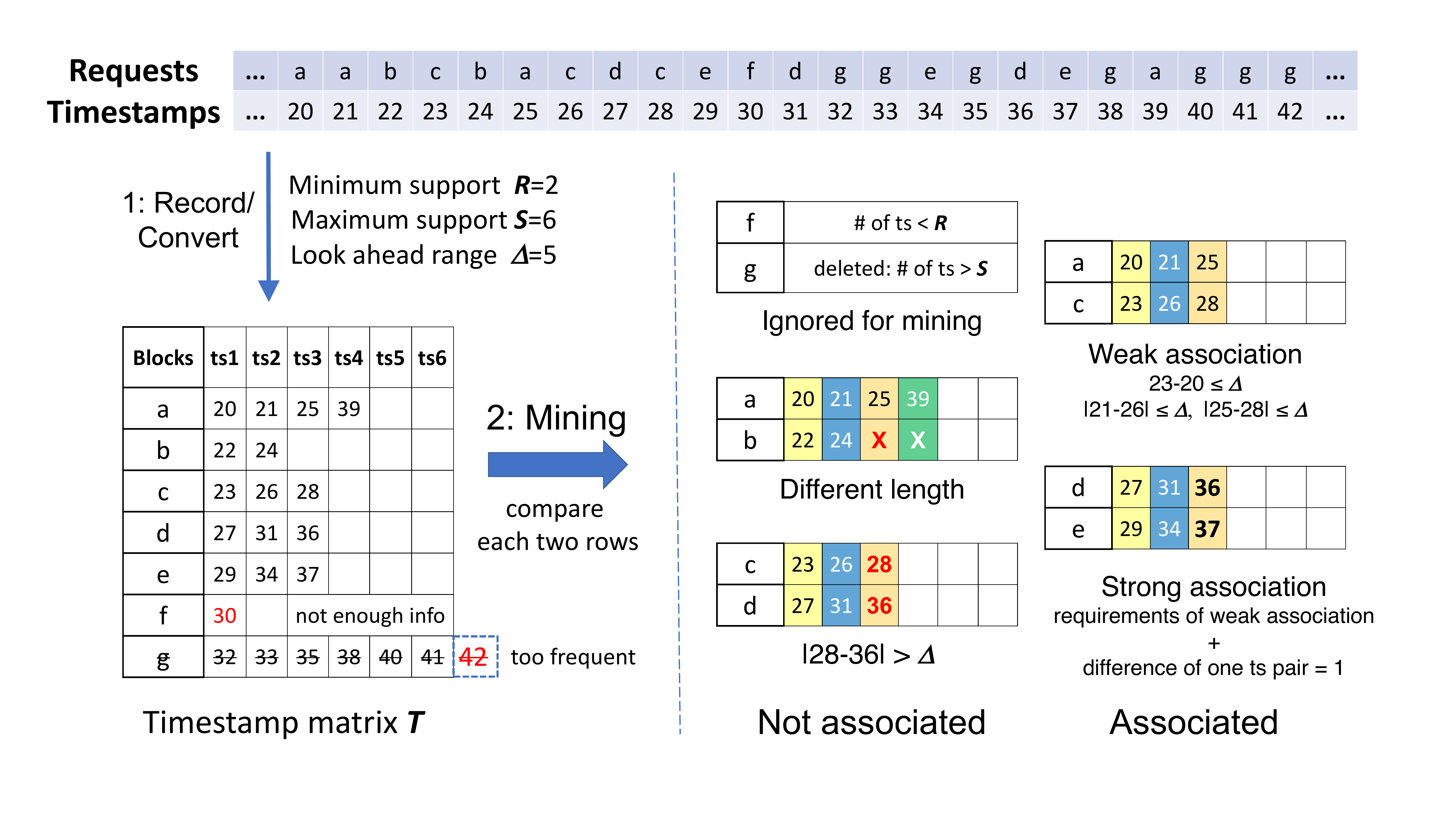}
\vskip-0.2in
\caption{\textbf{Illustration of mining procedure.} \textit{If input is a request sequence, convert it into time-stamp matrix \textbf{$T$}. Blocks that have fewer than \textbf{$R$} time-stamps(ts) or more than \textbf{$S$} time-stamps are not considered for mining. For each two-block pair, if they have  different numbers of time-stamps, or the difference between at least one time-stamp pair is greater than \textbf{$\Delta$}, they are not associated. If all time-stamp pairs are within $\Delta$, they are weakly-associated. Furthermore if they have at least one time-stamp pair with difference 1, they are strongly-associated. }}
\label{fig: mining}
\end{center}
\vskip-0.3in
\end{figure*}

We now describe the algorithm at the core of our prefetching layer. 
Let $B$ be a sequence of unique block I/O addresses $B = \{ b_1, b_2, \dots, b_n \}$ where a request $b_i$ has a logical \textit{time-stamp} of $i$, also known as its reference number. Let $T$ be an $n \times S$ matrix for $S=\textit{maximum support}$, where $ith$ row $\vec T_i$ corresponds to request $b_i$, and the cells of each row contain a sorted list of increasing time-stamps. In addition, $T$ is also sorted by the first time-stamp of each block. Figure~\ref{fig: mining} illustrates the request sequence and corresponding time-stamp matrix $T$ (all the symbols are listed in Table~\ref{tableSymbol}).

\begin{table}[!ht]
\centering
\caption{Symbols used in the text}
\label{tableSymbol}
\scalebox{0.8}{
\begin{tabular}{|l|l|}
\hline
Symbol     & Meaning                        \\  \hline
$T$ & \textit{Time-stamp Matrix}      \\  \hline
$R$ & \textit{Minimum Support}       \\  \hline
$S$ & \textit{Maximum Support}       \\  \hline
$\Delta$ & \textit{Lookahead range}       \\   \hline
$M$ & \textit{Maximum Metadata Size} \\  \hline
$P$ & \textit{Prefetching List Size} \\  \hline
\end{tabular}
}
\end{table}

An \textit{associated block pair} refers to two blocks that are \textit{repeatedly} accessed \textit{in sequence}. In modern systems, due to multiple applications interleaving with each other, two consecutive accesses from the same stream may not appear consecutive in the final stream, so we define a \lookaheadWindow{} that specifies the maximum allowed distance between two \textit{associated} blocks. In order to establish an association between two blocks, not only do they need to appear within $\Delta$ of each other, but also they need to appear with some minimum frequency. We denote this threshold as \minSup{}. 
Since our prefetching layer assumes the presence of a cache to catch frequent items, we specify \maxSup{} as the upper bound for items to be considered for mining within a certain time interval. We remark that each of these requirements have conceptual counterparts in sporadic association rule mining. 

To further distinguish associated block pairs, as illustrated in Fig~\ref{fig: mining}, we define two blocks as being \textbf{\textit{weakly associated}} if each time-stamp pair of the two blocks is within $\Delta$; furthermore, if a \textit{weakly associated} pair is accessed strictly consecutively (time-stamp difference 1) at least once, we define it as being \textbf{\textit{strongly associated}}. 

The reason for distinguishing \textit{weakly associated} pairs and \textit{strongly associated} pairs is that two blocks in a \textit{strongly-associated} pair are more likely to be related, which is preferred for prefetching. However, due to multiple applications interleaving, a \textit{strong association} does not always exist for each block, while there might be multiple \textit{weakly associated} pairs. Therefore, only a \textit{strongly associated} pair and the closest \textit{weakly associated} pair are considered. 

We present the basic version of \alg{} in Algorithm~\ref{algo: basic}. 
The function $checkAssociation$ (Algorithm~\ref{algo: checkAssociation}) receives two rows from $T$ as input and checks whether the corresponding two blocks are weakly or strongly associated or not. 

Algorithm~\ref{algo: basic} shows the mining procedure, which uses O($N$) time to discover associated block pairs. $N$ is the number of unique blocks requested during the recording interval. The input of the algorithm can be the request sequence $B$ or the time-stamp matrix $T$. If the input is $B$, then we need to first convert it into $T$ in O($N$) time. 

In the outer loop, we iterate through all rows in $T$. For each block $b_i$, we check all other blocks in the inner loop to find $b_j$ that are either \textit{strongly associated} or are the first \textit{weakly associated} occurrence. 
Because $T$ is sorted by first time-stamp of each block, so at inner loop at most $\Delta$ blocks are checked. 
Typically, the number of blocks checked is much less than $\Delta$. 

After an associated block pair is unveiled, it is stored in the \textit{prefetching table}, which is checked for prefetching upon each request.

\begin{algorithm}[!htb]	
\DontPrintSemicolon 
\KwIn{Rows $R1$ and $R2$ from time-stamp matrix $T$, associationType $assoc$, \lookaheadWindow{}}
\KwResult{Whether $b_1$ and $b_2$ are associated}

    $consecutive \gets False$\; 
	\If{$len(R1) - len(R2) \neq 0$}{
		\Return{False}\;
	}
	\For{$k \gets 1$ \emph{\KwTo} $len(R1)$}{
		\If{$abs(R1[k] - R2[k]) > \Delta$}{
    		\Return{False}\;
		}
		\If{$abs(R1[k] - R2[k]) == 1$}{
			$consecutive \gets True$\;		
		}
	}
\If{$assoc == weak$} {
    \Return{True}\;
}
\ElseIf{$assoc == strong$}{
    \Return{$consecutive$}\;
}

\caption{checkAssociation}
\label{algo: checkAssociation}
\end{algorithm}

\begin{algorithm}[!htb]	
\DontPrintSemicolon 
\KwIn{time-stamp matrix $T$, \minSup{}, \lookaheadWindow{}}  
\KwResult{Associated block pairs}
\For{i=1 \emph{\KwTo} len(T)-1}
{   
    \If{$len(T[i]) < R$}{
        continue\;
    }
    $associationType \gets weak$\;
    \For{j=i+1 \emph{\KwTo} len(T)}
    {
        \If{checkAssociation($T$[$i$], $T$[$j$], $associationType$)}{
            $addAssociation(block_i$, $block_j$)\;
            $associationType \gets strong$\; 
        }
        \If{$ (T[j][0] - T[i][0]) > \Delta $}{
            break\;
        }
    }

}

\caption{\alg{} mining procedure}
\label{algo: basic}
\end{algorithm}

\subsection{Optimizations}
When \alg{} is run, a two-dimensional time-stamp matrix $T$ is initialized. For each new request, if it is found in $T$, the current time-stamp is appended to the corresponding row. Otherwise, the request is recorded in a new row. We append the time-stamp to a row. When the row is full,  the block is considered frequent and deleted from the matrix and recorded in the frequent block hashmap. Items from this hashmap are ignored when encountered again before the mining process. When the time-stamp matrix $T$ is full, the mining procedure is called and the associated blocks are saved in the prefetching table. After mining completes, recording starts anew with a clean state. 


The version of \alg{} described so far requires a large matrix with \maxSup{} columns for storing time-stamps, a hashmap mapping from block number to the corresponding row in the matrix and a hashmap for determining whether a block is frequent. Additionally, a prefetching table is needed for storing associated block pairs for prefetching. However, spending limited cache space on tracking large metadata is not desirable. 
To address the metadata space usage of basic \alg{}, we made the following optimizations, which use bounded memory in exchange for some added complexity. 

\subsubsection{Recording and Mining}

{\textbf{Splitting recording table.}} The two-dimensional recording table (time-stamp matrix) is a sparse matrix, since a typical block, by definition, will be requested fewer than \maxSup{} times within a recording period. 
A na{\"i}ve implementation uses a linked list for each block instead of a fixed-size array. However, the space for link pointers between timestamp nodes doubles the space overhead. 
We exploit the sparsity by decomposing the large matrix into two smaller fixed-sized tables: one with \minSup{} columns, which is the \rt{}, and the other one with \maxSup{} columns, which we call the \mt{}.
The \rt{} is a circular array in which new entries replace old entries in FIFO fashion. The \mt{} is a fixed-size array that triggers the mining procedure when full. 

When a block request arrives, the time-stamp is recorded in the \rt{}. If the number of time-stamps in the corresponding row has reached \minSup{}, in other words, when the row is full, it is declared to be \textbf{mining ready} and then transferred into the \mt{}, which can store up to $S$ time-stamps for each block. After migrating one row from the \rt{} to the \mt{}, the last row in the \rt{} is moved up to the migrated row to make the table compact. When the \mt{} is full, the mining procedure is triggered to discover associated block pairs and store them in the \pt{} for prefetching. When the mining finishes, the \mt{} is cleared. When the \rt{} is full, we replace the oldest entry with a new entry with the assumption that the oldest block remaining in the table is rare since it has not been requested $R$ times within the interval. 

Decomposing the original matrix not only saves space but also allows for more blocks to be tracked. Because the \rt{} does not need to be cleared each time, we retain extra information for blocks that are not mining ready. In the unoptimized approach, the large time-stamp matrix was cleared each time the mining finishes, discarding all information. 

The primary drawback of splitting is that the \mt{} needs to be sorted before mining. This is because Algorithm~\ref{algo: basic} requires input to be sorted by the first time-stamp, which occurs automatically in our single-table construction. Since our separate \mt{} is created by inserting elements in the order of accumulating $R$ time-stamps, sorting the \mt{} before mining is necessary. 
In practice, however, the \mt{} is usually tiny and sorting is trivial. 


{\textbf{Compressing time-stamps.}}
To further reduce the space used by the \rt{} and the \mt{}, we compress time-stamps by storing only the lower 15 bits. This allows us to store four time-stamps in the lower 60 bits of one 64-bit integer with a time-stamp counter stored in the higher 4 bits. Moreover, one could further compress time-stamps by removing the last $\floor{\log_2(\Delta)}$ bits -- we omitted this optimization in our experiments to limit time overhead. 

{\textbf{Removing the frequent block hashmap.}}
A block that is requested more than $S$ times in each recording interval in the original \alg{} approach is considered to be a frequent block, so no information should be recorded. To track the requests, one could use a hashmap or Bloom filter, but hashmaps require extra memory and Bloom filters incur extra computation overhead. Instead, we decide to record a block only on cache miss. In this way, all frequent blocks are automatically filtered out by the underlying cache. There are several other benefits. 
First, \alg{} need not be invoked when cache sizes are sufficiently large and \minSup{} is greater than 1. This behavior happens gradually over larger cache sizes since the mining phase will be run less frequently.
Second, if a block is accessed frequently over a short period, the optimized recording method cuts down overhead since  it only records cache misses, thus precluding spuriously recording frequently accessed blocks. If the cache size is small, recording bursts and thus prefetching frequent items is useful since these blocks are constantly being evicted by the underlying cache.


Our optimizations trade off storage, computation overhead, prediction precision and hit-ratio improvement. The more useful information we record, the higher hit rate and precision can be achieved, but at the same time more overhead is incurred. Besides recording at cache miss as mentioned above, optionally we can also record the time-stamp when a block is evicted from the cache to obtain more information about the block. Recording at eviction is similar to recording at cache miss: in both approaches, the frequent blocks are filtered out by the underlying cache. 
 

\subsubsection{Prefetching}

{\textbf{Splitting the prefetching table into shards.}} 
We use a two-dimensional array instead of lists to store associated block pairs together for storage reduction for the same reason as using an array in the \rt{}. 
In the \pt{}, the first column stores the originated block number $b_x$, while the rest of the columns store the blocks that are associated with $b_y$. The number of columns left is the maximum number of possible block pairs, defined as \prefetchingListSize{}. We use a default of three columns, indicating that, at most two block pairs can be stored for each block. For example, in an association $b_x \rightarrow b_y$, $b_x$ is stored in the first column and $b_y$ is stored in the second column. If there is another association, $b_x \rightarrow b_z$, then the third column stores $b_z$. If more than two associations are discovered, we replace the old associations in a FIFO manner, which allows \alg{} to adapt to changing workloads. 

Since cache behavior varies in different workloads, it is impossible to know how many blocks will have associations ahead of time, and thus how much memory will be needed. Therefore, we introduce the concept of shards. A shard is a \pt{} with 2000 rows that is dynamically allocated when needed. When a user specifies a \maxMetaDataSize{} can be used for \alg{}, an upper bound is placed on the number of possible shards. When all possible shards are allocated, a new row will replace the oldest row. 

By introducing shards, we aim to find a balance between frequent allocation and overallocation of memory. In addition to saving metadata memory usage, the maximum memory usage is also bounded by \maxMetaDataSize{}. 


Since prefetched blocks are also added to the original cache pool, it is possible for a prefetched block to be evicted before it is used. 
As other authors suggest~\cite{gill2005sarc, gill2007amp}, we give the prefetched block a second chance by re-adding it to the MRU end of cache if it is going to be evicted without being accessed.

\subsection{Using \alg{}}
Using \alg{} as a prefetching layer requires minor modifications to the underlying caching layer. The complete flow of \alg{} is shown in Algorithm~\ref{algo: complete}. A prefetch from \alg{} requires passing one parameter and two indicators. The parameter is the current block number, which is used for recording, prefetching or both. The two indicators are $\text{pFlag}$ and $\text{rFlag}$, which indicates whether it is for recording or prefetching. 

There are two scenarios where the \alg{} API may be called.
First, when a request arrives, \alg{} must check whether prefetching is needed. In this situation, $\texttt{pFlag}=\texttt{True}$ and $\texttt{rFlag}=\texttt{False}$. Second, to handle recording when $\texttt{rFlag}=\texttt{True}$ and $\texttt{pFlag}=\texttt{False}$. This recording may be invoked (a) at the arrival of each request, (b) only at cache misses, (c) only during cache eviction, or (d) during both misses and eviction. 
Recording at each request or recording at both misses and evictions increases the computation overhead. As we demonstrate in Section~\ref{sec: params}, recording on the arrival of each request optimizes performance, whereas recording only at cache misses provides similar performance at much lower overhead. In contrast, we find the two approaches (c, d) recording on eviction do not to provide competitive performance. 


\begin{algorithm}[!t]
\DontPrintSemicolon 
\KwIn{recording table $rTable$, mining table $mTable$, prefetching table $pTable$, minimum support $R$, block\# $b$, prefetchingFlag $pFlag$, recordingFlag $rFlag$}
\KwOut{blocks to prefetch} 
$ts \gets 0$\;

\If{$rFlag$}{
	$tsRow \gets pTable$[$b$] \;
	append $ts$ to $tsRow$ \;
	\If{$len(tsRow) \geq{} R $}{
		move $tsRow$ to $mTable$\;
		move last row in $rTable$ to $tsRow$\; 
		\If{ $mTable$ is full}{
			mining()\;
			clear $mTable$\; 
		}
	}
	$ts \gets ts + 1 $\;
}
\If{$pFlag$}{
	\If{$b$ $\mathbf{in}$ $pTable$}{
		\Return{$pTable$[$b$]}\;
	}
}
\Return{NULL (no need to prefetch)} 
\caption{Description of the \alg{} algo.}
\label{algo: complete}
\end{algorithm}


\subsection{Complexity Analysis}
\textbf{Time complexity.} Compared to \LRU{}, the only operations added to each request are to record the current logical time-stamps in the \rt{} on a cache miss and check the \pt{} and prefetch when needed. Each of these operations has a time complexity of $O(1)$, so the total computation overhead at each request is negligible. Periodically, the mining procedure runs and is dominated by an $O(N\log N)$ sort, where $N$ is a fixed, typically small table size. The mining process can furthermore be run in a background thread and thus avoid blocking new requests. 


\textbf{Space complexity.} In the optimized \alg{}, we store all time-stamps as 15-bit integers with four time-stamps in one 64-bit integer. Thus if we have \maxSup=8, \minSup=4, \rt{} size 100,000 and \mt{} size 1,250, recording and mining will need
less than 2MiB.
When calculating size of \texttt{hashtable}, which maps from block address to index in \rt{} or \mt{}, the 8 byte is used for storing block address, the 4 bytes is used for storing the index. 

Since all information is stored in a bounded array, the \maxMetaDataSize{} allocated is usually set to 10\%, which is more than enough in most cases. 
And in our evaluation, we count in the memory usage for all metadata for fair comparison, which means when \alg{} metadata uses 5\% of cache space, then only 95\% of space will be used for store cache data.

\section{Evaluation} 
\label{sec:eval}


We now characterize \alg{} experimentally with the following questions in mind: 
\begin{itemtight}
\item How much does \alg{} improve the hit ratio? What are the best and worst cases? 
\item How well does \alg{} work with various cache replacement algorithms, and how precise is prefetching? 
\item How do parameters affect \alg{}? 
\item Is latency improvement enough to justify overhead? 
\item Why does \alg{} work? 
\end{itemtight}

\subsection{Methodology}
\label{sec:methodology}
As a history-based prefetching layer, ideally we should compare \alg{} with C-Miner \cite{li2004cminer} and QuickMine \cite{soundararajan2008quickmine}, which are the two state-of-the-art algorithms in history-based prefetching. However, since  C-Miner and QuickMine either runs offline or requires context information from the application, which is not applicable in our setting. 
So instead we implemented another history-based prefetching technique, \PG{} (\PGs{}) \cite{griffioen1994probabilityGraph}, together with a state-of-the-art sequential prefetching algorithm, \amp{} \cite{gill2007amp}, and \LRU{} to compare to \alg{}. Note that \alg{} can be used on top of \amp{}. 

We evaluated algorithms on 106 traces from commercial I/O caching analytics services from CloudPhysics (CP)  \cite{Waldspurger2015Shards} together with 29 traces obtained by Microsoft Research (MSR) \cite{Narayanan2008MSR} (We omitted  traces that have fewer than a half million requests). For simulation-based results, we used the \textsc{mimircache} \cite{mimircache} for profiling and analysis on a Microway server of dual E5-2670v3 CPUs with 512GB memory. For the micro benchmark, we modified IOBlazer \cite{ioblazer} and ran it on AWS EC2 c3.large instance with an EBS magnetic disk. In this section, if not specified, \alg{} is used together with \LRU{}, and all experiments showing single trace used trace w94 from CP \cite{Waldspurger2015Shards}, which is a week-long VM trace. The cache size, if not mentioned, is set to 256MB, which exhibits a range of \LRU{} hit ratios between 10\% to 99\%. The profiling platform and \alg{} implementation will be released under open-source after publication \cite{mimircache}. The CP data used in the paper will be released by CloudPhysics separately.

\begin{figure*}[!t]
  \centering
  \vskip-0.3in
  \includegraphics[height=0.96\textwidth, angle=-90]{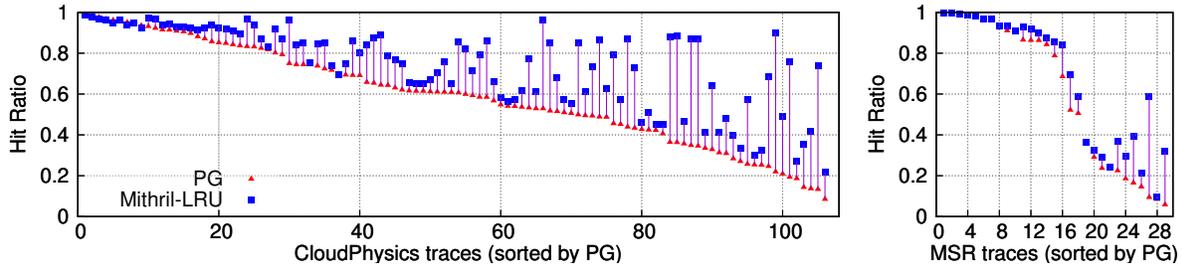} 
  \caption{\textbf{Hit ratio of \PGs{} and \alg{} for 106 CP traces and 29 MSR traces sorted by \PGs{} hit ratio.} \textit{Hit ratio of \LRU{} omitted as it is similar to \PGs{} (Pearson $r=0.995$ compared to $r=0.742$ for \LRU{} and \alg{}). Compared to \PGs{}, \alg{} overall provides significant improvement, even though parameters are not fine-tuned for each trace. }} 
  \label{fig:hr_LRU}
    \vspace{-2em}
\end{figure*}



\begin{figure*}[!t]
\centering
\begin{subfigure}{0.48\linewidth}
  \centering
  \includegraphics[height=\linewidth, angle=-90]{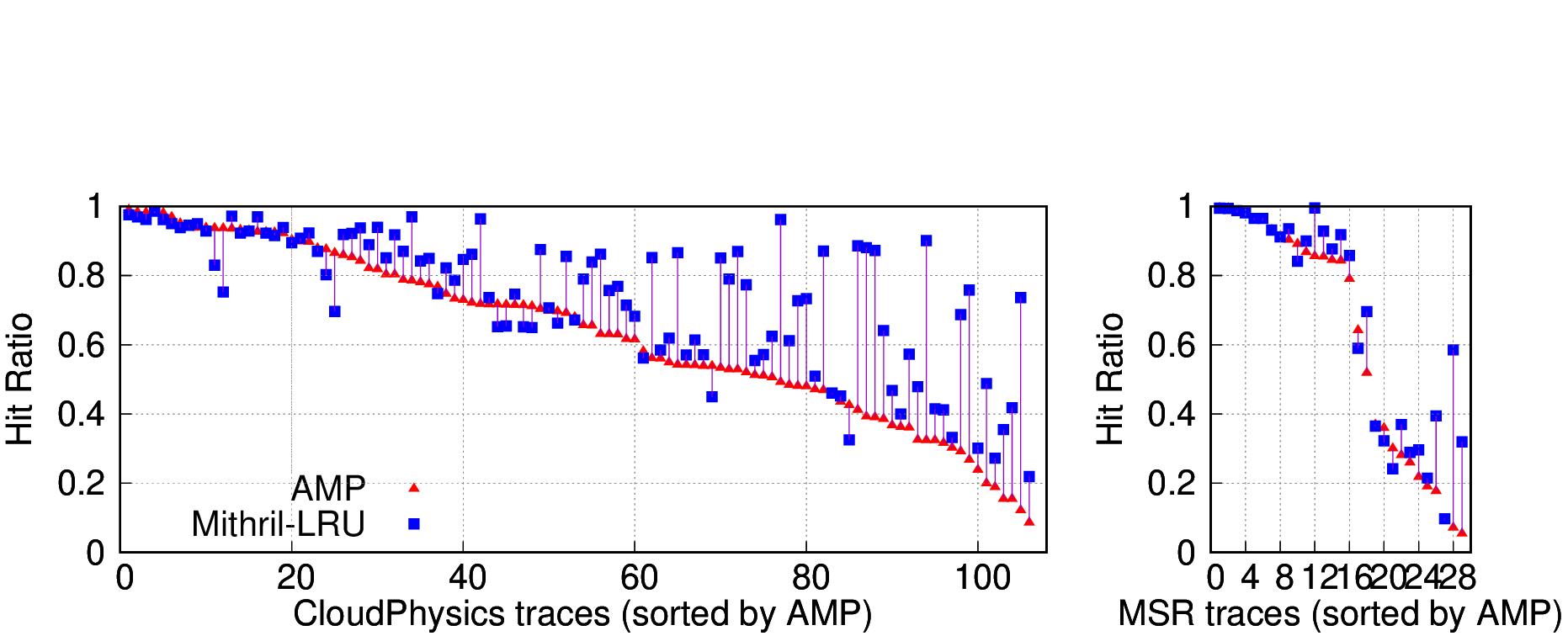} 
\end{subfigure}
\begin{subfigure}{0.48\linewidth}
    \centering
  \includegraphics[height=\linewidth, angle=-90]{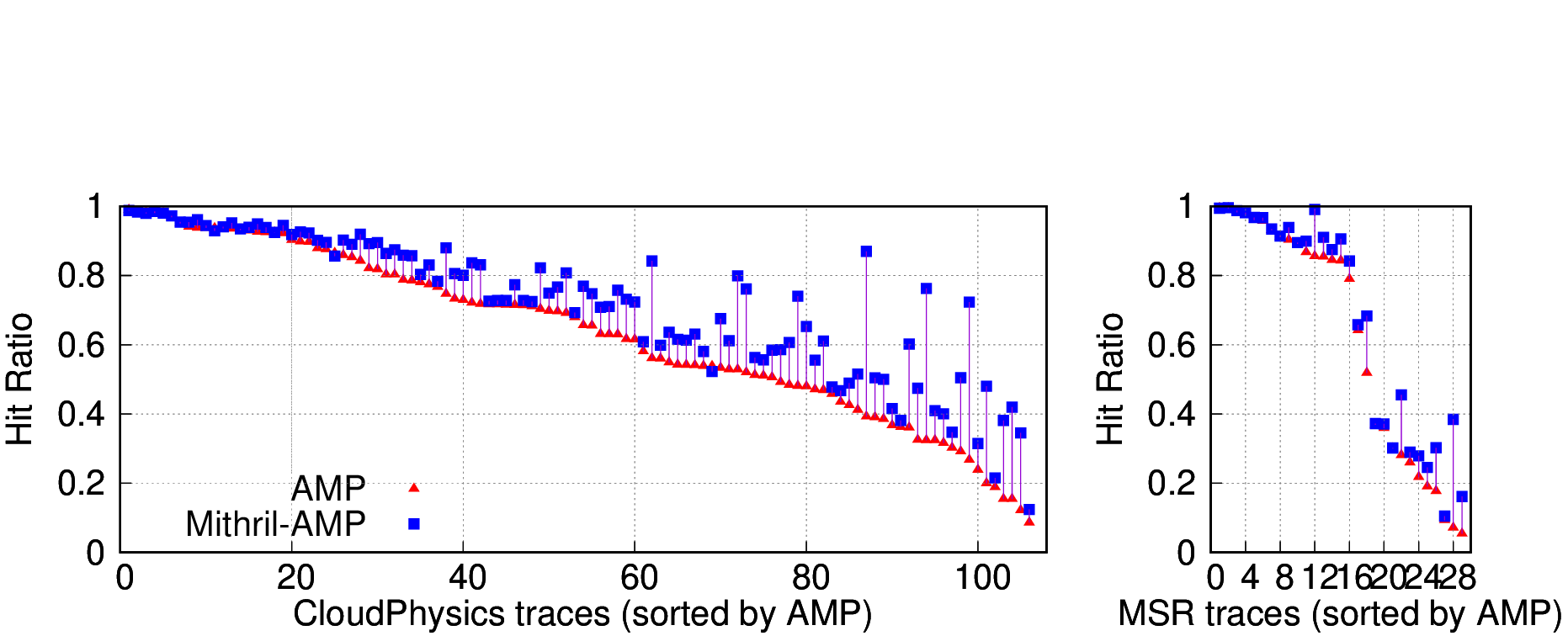} 
\end{subfigure}
  \caption{\textbf{Left: Hit ratio of \amp{} and \mLRU{}, right: Hit ratio of \amp{} and \mAmp{} for CP and MSR traces sorted by \amp{} hit ratio.} 
  \textit{Left: \mLRU{} outperforms \amp{} in most traces. For some traces with strong sequentiality, \amp{} has better performance due to its ability to prefetch pages that have never been requested. Right: \mAmp{} improves or matches hit ratio for most traces compared to \amp{}.}} 
  \label{fig:hr_amp0}
  \label{fig:hr_amp1}
\vspace{-0.8 em}
\end{figure*}

\subsection{Overall Hit Ratio Improvement}
\label{sec: hitRate}
As a prefetching layer unaware of the underlying caching layer, which can be either FIFO, \LRU{}, \amp{} or any other cache replacement algorithms, In this section, we show that \alg{} provides benefits for all of them. 

\textbf{Comparison with PG.}
\PGs{} is the most comparable history-based algorithm, so we compare \alg{} with \PGs{} in this section. In Figure~\ref{fig:hr_LRU}, we show the hit ratio of \PGs{} and \alg{} for all the traces. \LRU{} is not shown in the trace because of its high resemblance to \PGs{} in terms of average hit ratio and correlation: the Pearson Correlation Coefficient between hit ratio of \LRU{} and \PGs{} is 0.993, while it is 0.801 between \LRU{} and \alg{}. The low correlation between \LRU{} and \alg{} implies that the performance of \alg{} does not completely depend on the performance of \LRU{}.
Compared to \LRU{}, on average \alg{} provides 52\% relative improvement in the hit ratio on CP traces, and on average achieves 82\% of the maximum obtainable hit ratio at small cache size, which is calculated by excluding cold miss. 
On the 29 MSR traces, \alg{} provides on average a 64\% hit ratio improvement achieving 81\% of the maximum obtainable hit ratio. 
As shown in the figure, the hit ratio improvement for \alg{} varies between traces. For certain traces, it can provide up to more than 7$\times$ improvement, but for some other traces, the improvement is more modest, particularly those whose \PGs{} hit ratio is already high. 

\textbf{Comparison with \Amp{}.}
As a prefetching layer, we also compare \alg{} with state-of-the-art sequential prefetching algorithm \amp{}, which dynamically captures the spatial associations in the requests. 
Compared to \amp{}, \alg{} on average provides a 31\% increase in hit ratio on CP traces and 51\% on MSR traces, indicating that by exploring temporal associations, \alg{} can provide more benefit than \amp{}. 
However, as shown in Figure~\ref{fig:hr_amp0}, \alg{} does not always provide more benefit compared to \amp{}. In some traces where sequentiality is not dominant, \alg{} provides a great benefit, more than a 7$\times$ improvement on hit ratio; in some other traces where sequentiality dominates the disk access pattern, \amp{} provides more benefit than \alg{}. The reason \amp{} outperforms \alg{} lies in its ability to prefetch blocks that have never been requested. In contrast, \alg{} does not have this ability. It can only prefetch blocks already seen in the past. 

Although \amp{} surpasses \alg{} in some cases, \alg{} as a prefetching layer can be used on top of \amp{}. In Figure~\ref{fig:hr_amp1}, we show the hit ratio obtained by \amp{} compared to \mAmp{}. Using \alg{} on top of \amp{} guarantees at least similar performance as \amp{}, and still provides a large benefit on most of the traces. This improvement implies that besides spatial-locality, which has been captured by \amp{}, \alg{} is capable of further leveraging the temporal-locality associations between requests to gain performance promotion. Note that Figure~\ref{fig:hr_LRU} and Figure~\ref{fig:hr_amp0} cannot be directly compared, because former one is sorted by \PGs{}, and latter one is sorted by \amp{}. However, Figure~\ref{fig:hr_amp0} and Figure~\ref{fig:hr_amp1} are comparable since curves in both figures are sorted by the \amp{} hit ratios. 
Adding \alg{} to \amp{} guarantees no performance loss compared to \amp{}, however, \mAmp{} does not guarantee a better performance than \mLRU{} as we see in some of the traces. The reason \mLRU{} can be better than \mAmp{} is that \amp{} turns some cache misses into cache hits due to its sequential prefetching ability. Thus the relationship seen by \alg{} is jeopardized, and the associations captured by \alg{} can be inaccurate. 
Overall, \alg{} significantly improves hit ratio over \PGs{} and \amp{}.



\begin{figure}[!t]
\begin{subfigure}{0.495\columnwidth}
  \centering
  \includegraphics[width=0.99\linewidth]{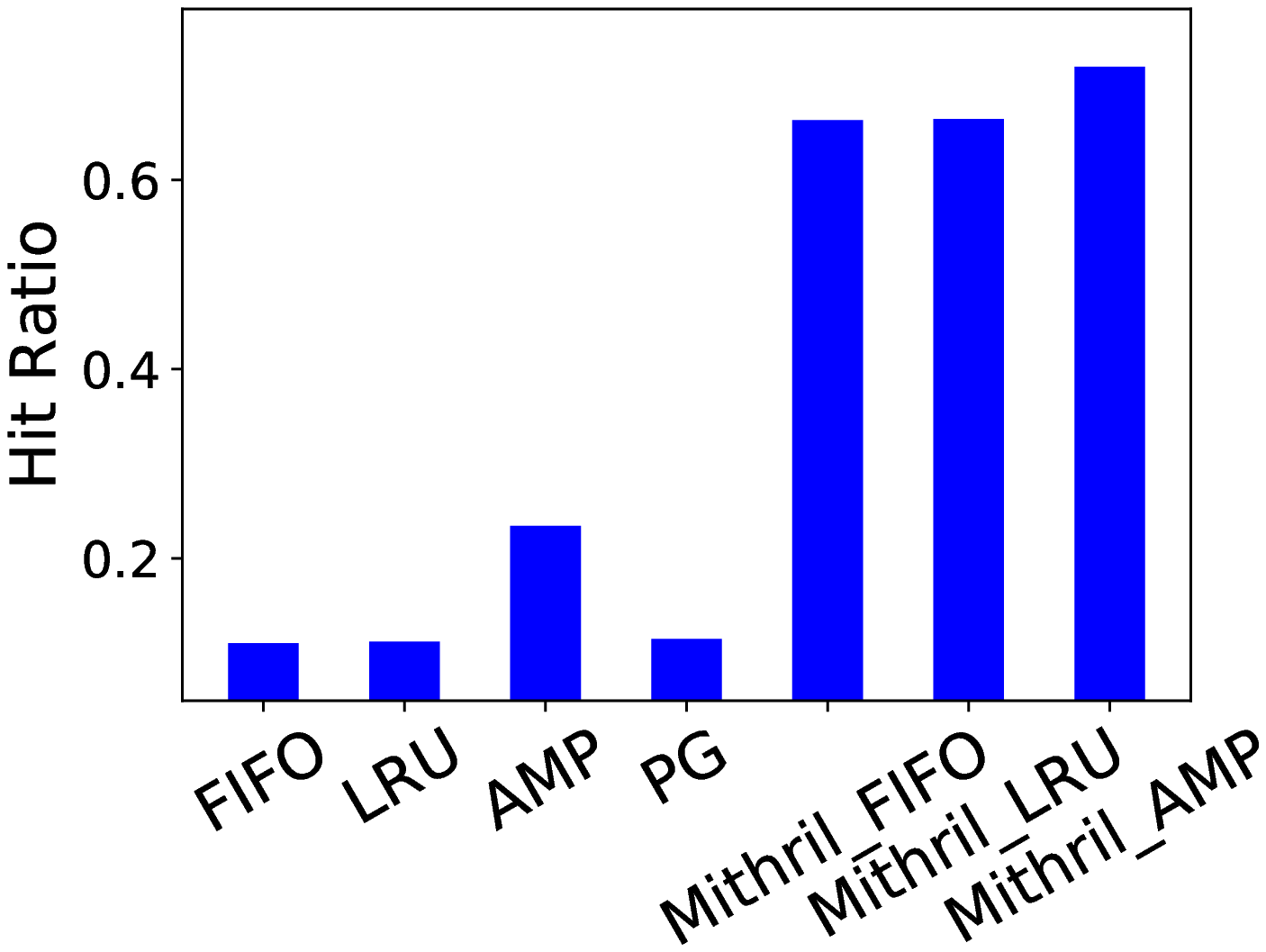} 
  \caption{CP trace w94}
  \label{fig: w94}
\end{subfigure}
\begin{subfigure}{0.495\columnwidth}
  \centering
  \includegraphics[width=.99\linewidth]{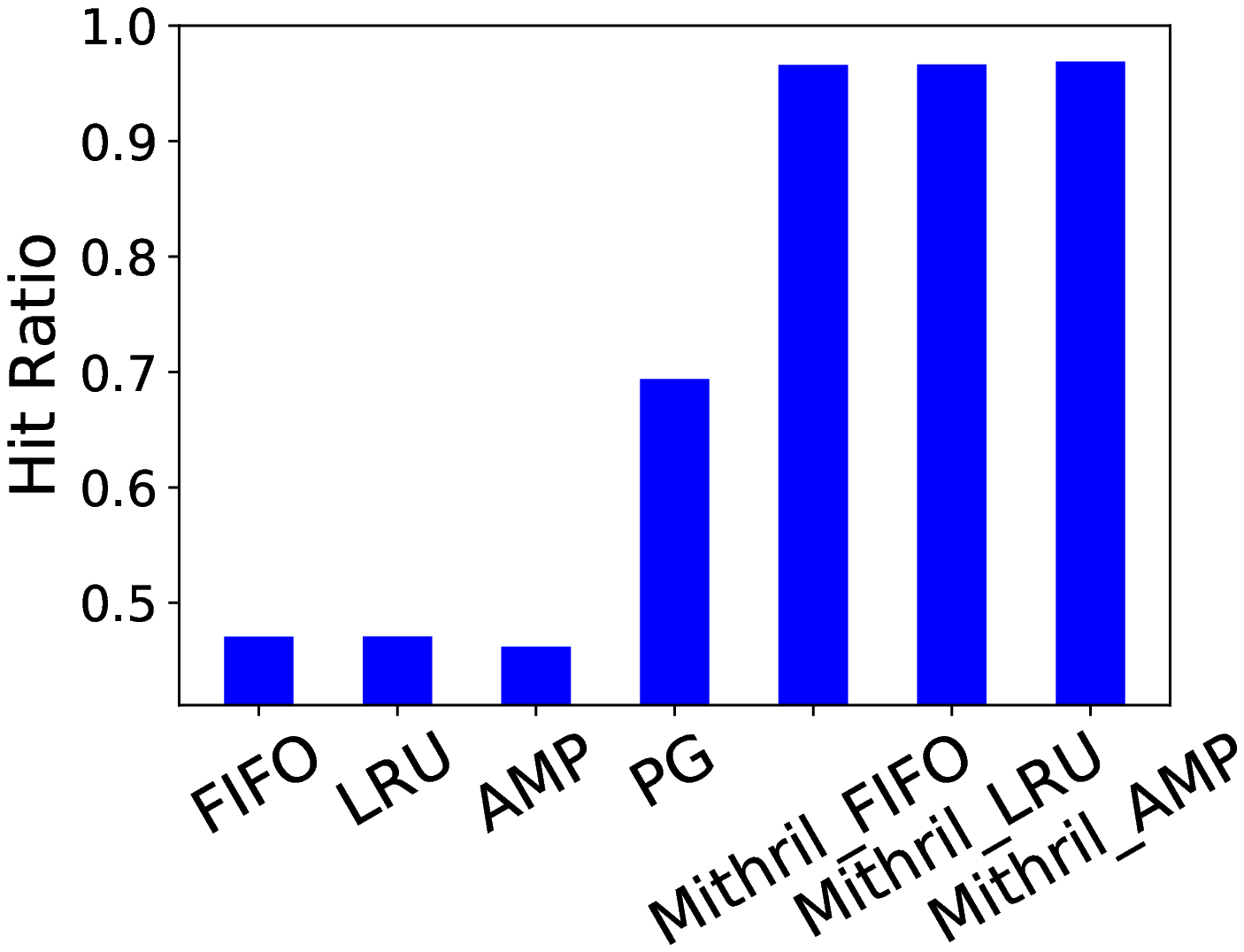} 
  \caption{MSR trace prxy}
  \label{fig: prxy}
\end{subfigure}
\begin{subfigure}{0.495\columnwidth}
  \centering
  \includegraphics[width=.99\linewidth]{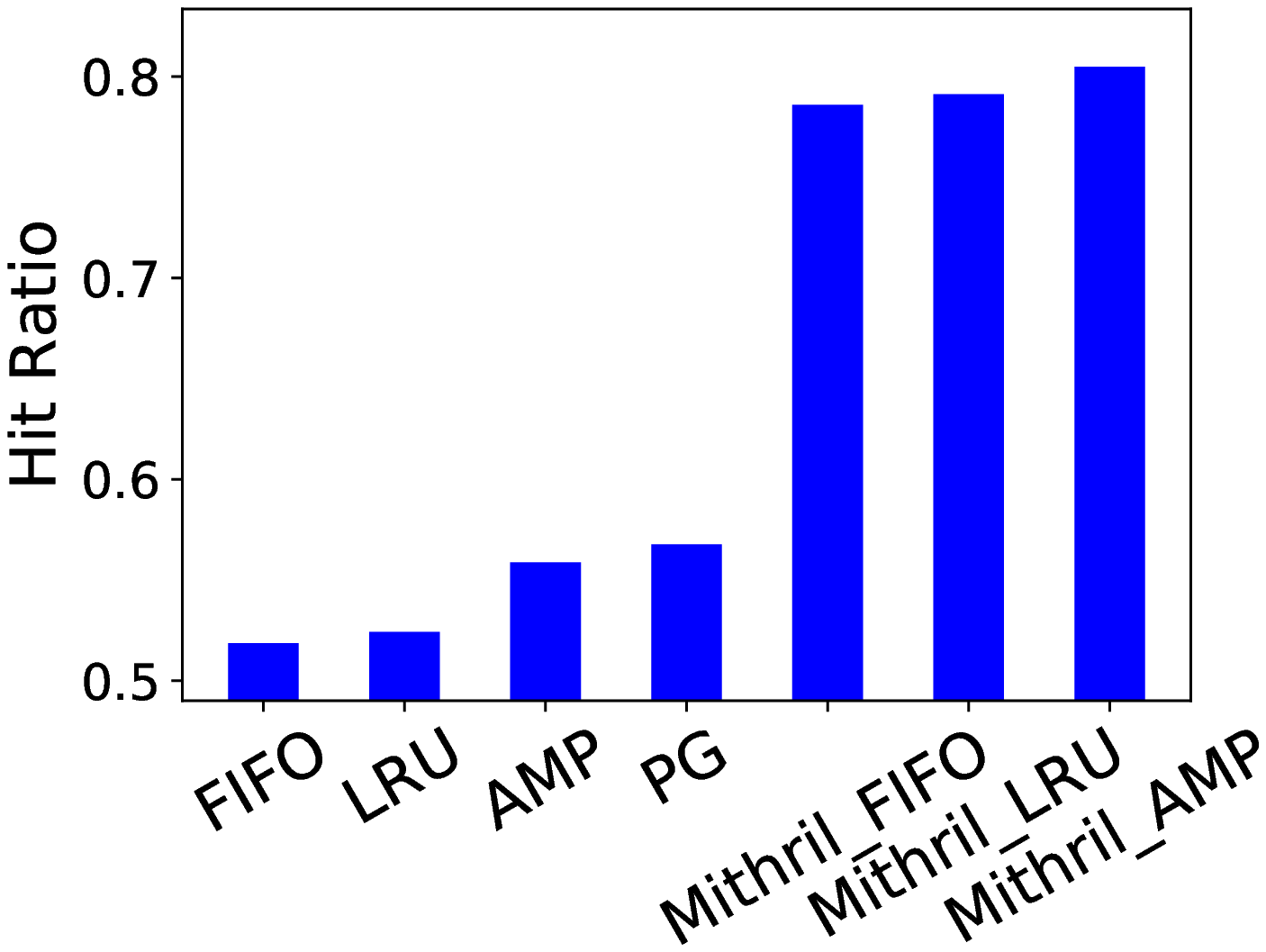} 
  \caption{CP trace w91}
  \label{fig: w91}
\end{subfigure}
\begin{subfigure}{0.495\columnwidth}
  \centering
  \includegraphics[width=.99\linewidth]{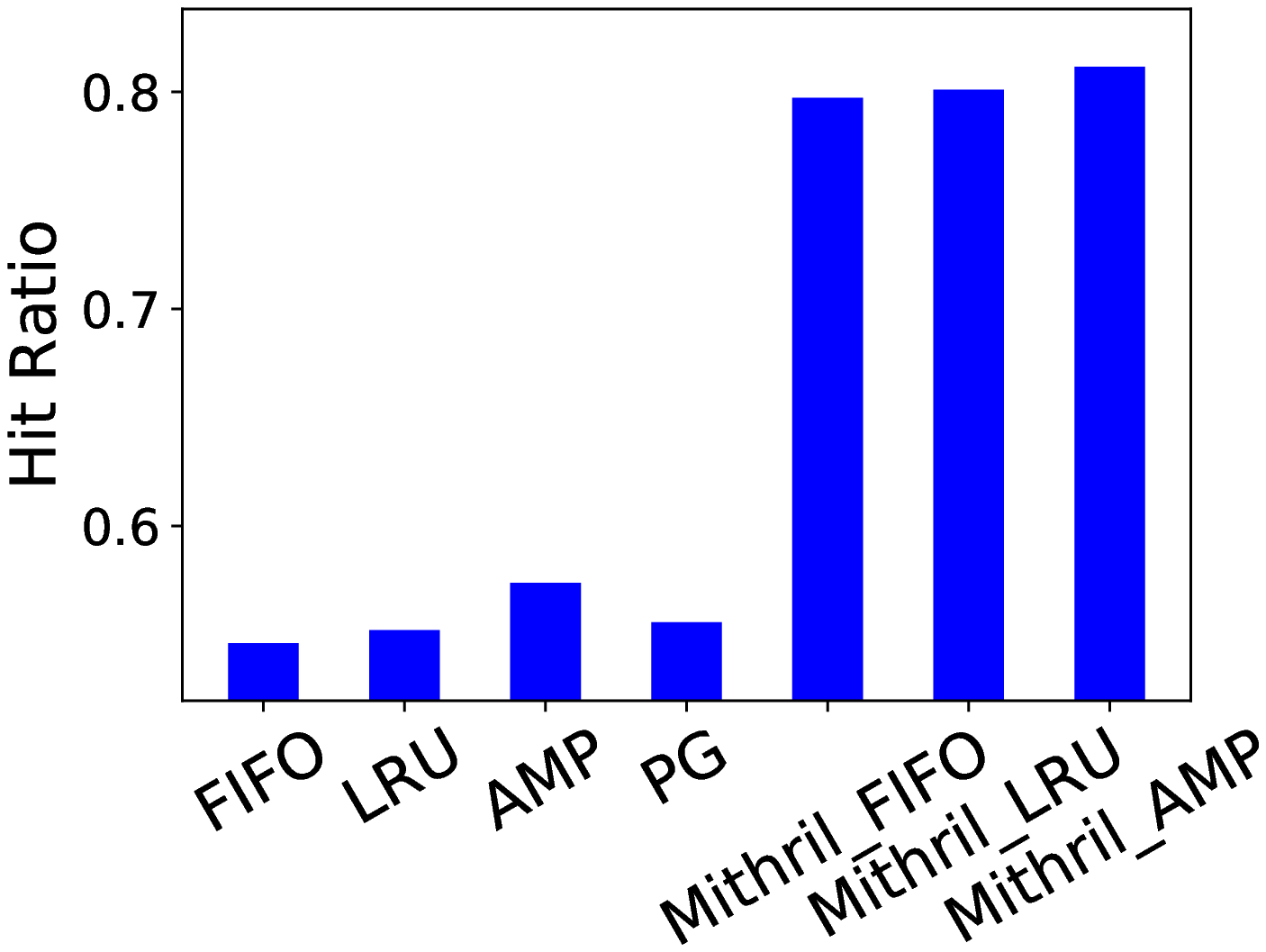} 
  \caption{MSR trace src1}
  \label{fig: src1}
\end{subfigure}
\begin{subfigure}{0.495\columnwidth}
  \centering
  \includegraphics[width=.99\linewidth]{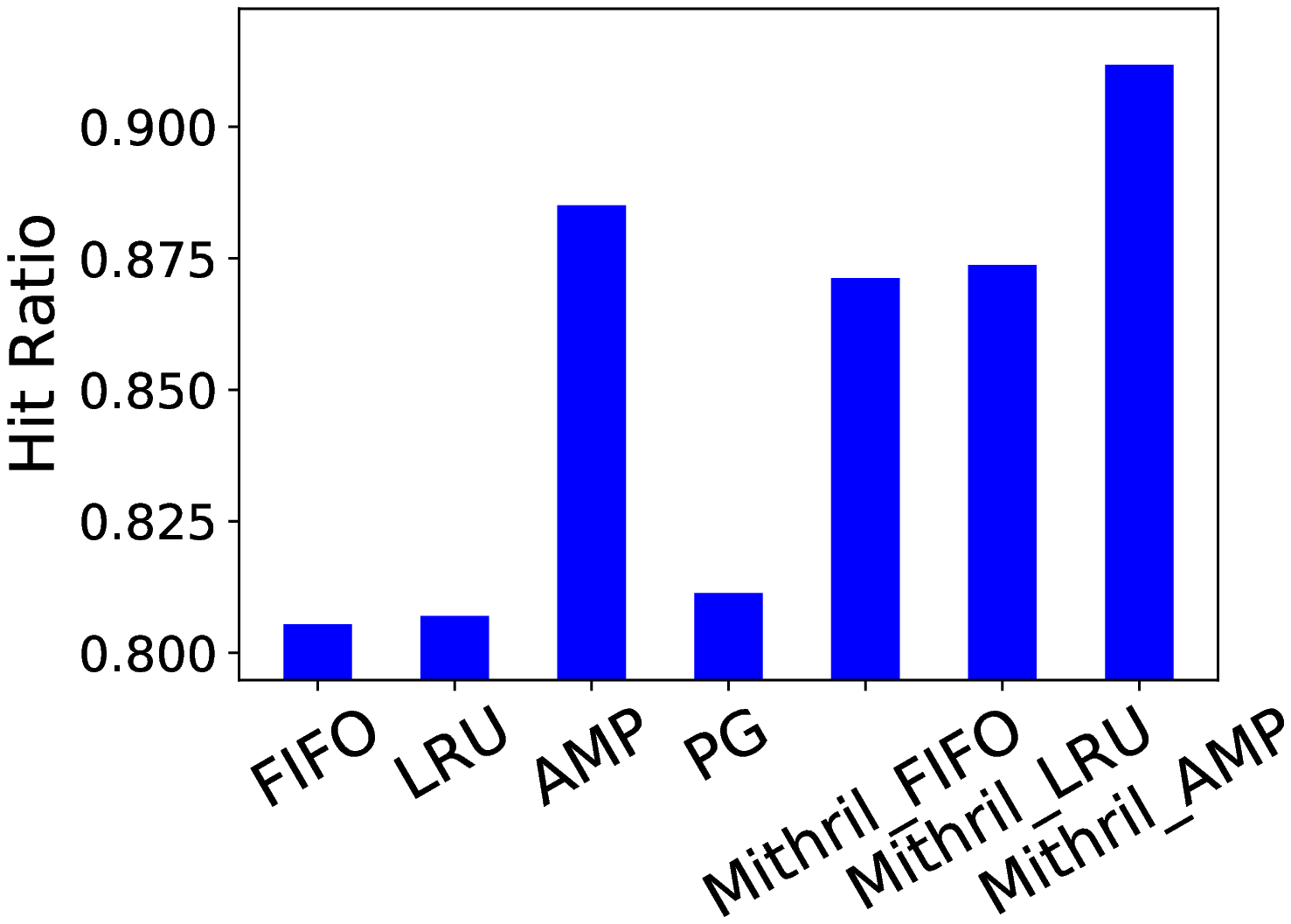} 
  \caption{CP trace w89}
  \label{fig: w89}
\end{subfigure}
\begin{subfigure}{0.495\columnwidth}
  \centering
  \includegraphics[width=0.99\linewidth]{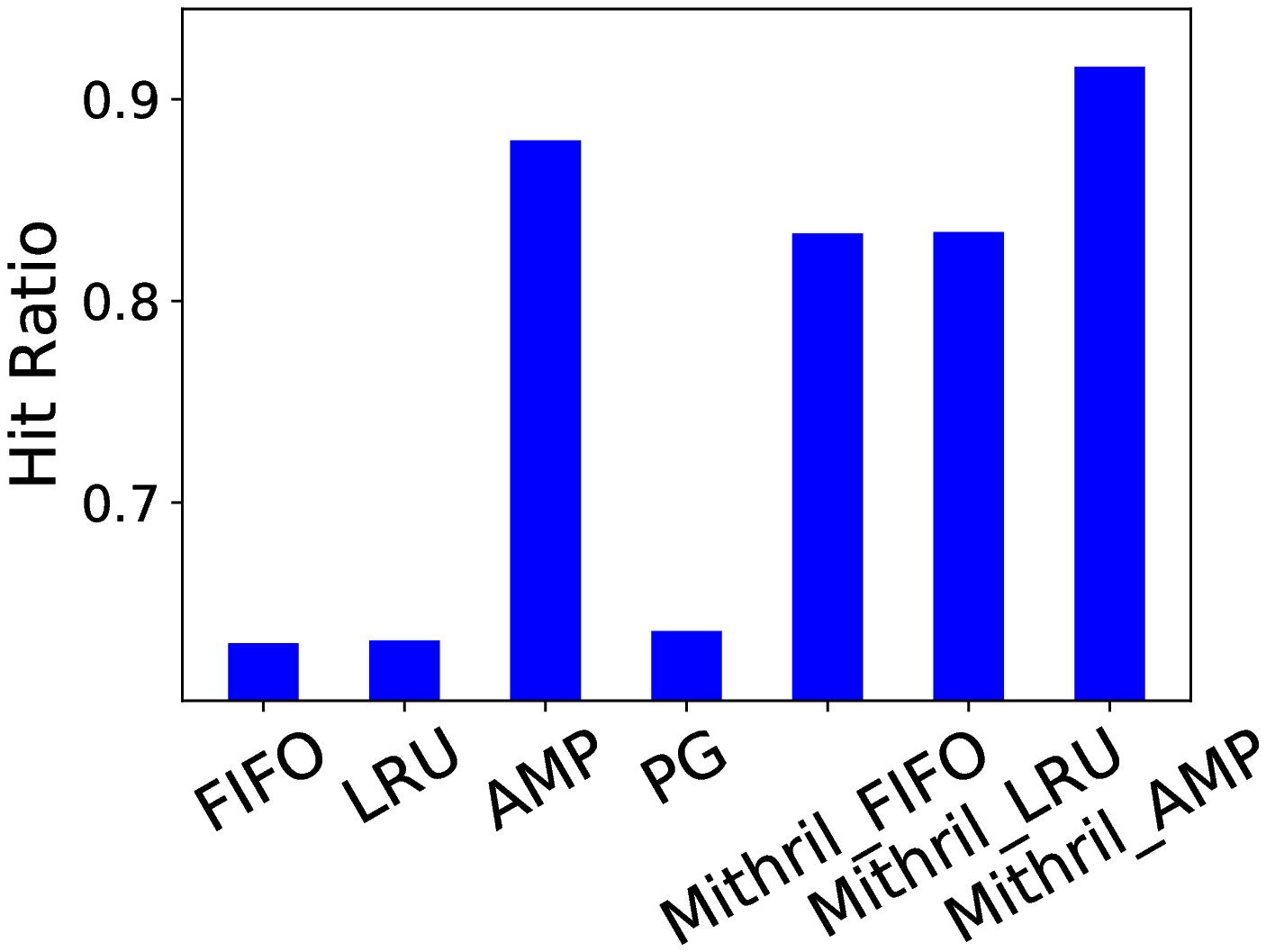} 
  \caption{MSR trace proj}
  \label{fig: proj}
\end{subfigure}

\caption{\textbf{Hit ratio of different algorithms.} \textit{Example traces where \alg{} significantly improves hit rate (top two), where \alg{} shows modest improvement (middle two), and where \alg{} shows little or no performance gain (bottom two). }} 
\label{fig: hr_alg}
\end{figure}

\textbf{Behavior on representative traces.}
To better illustrate the hit ratio improvement, we select six traces (three from CP and three from MSR) to show typical examples of  large (top two), modest (middle two) and small (bottom two) performance gains for \alg{} in Figure~\ref{fig: hr_alg}. 
The top two traces show the cases where \alg{} outperforms the corresponding caching algorithm by at least doubling the hit ratio. The middle two figures show the traces that have relatively high hit ratios under \LRU{}. Adding \alg{} provides a modest performance improvement. In the bottom two traces, \amp{} outperforms \mLRU{} by being able to prefetch unseen blocks. However, this can be changed by using \alg{} with \amp{}. Still, in these cases, \mAmp{} usually does not win over \amp{} much in terms of hit ratio because the hit ratios are often already high, limiting potential benefit. \alg{} can also only prefetch  blocks that have  already been seen, capping the maximum hit rate at $1 - \textit{cold miss ratio}$. 
\PGs{} is the only prefetching algorithm in same category as \alg{}. Its performance is unstable, sometimes better than \amp{}, most of time worse than \amp{}. For most traces, it outperforms pure \LRU{} and is outdone by \alg{}.

\alg{} is compatible with a range of caching algorithms. The figures compare performance of using \alg{} on top of \LRU{}, \FIFO{} and \amp{} to that of the original cache replacement algorithms. Adding \alg{} consistently boosts hit ratio, particularly for simpler cache replacement algorithms. For example, by adding \alg{} to \FIFO{}, the performance of \textsc{Mithril-FIFO} is similar to \mLRU{}, which is much better than \FIFO{}. This property of \alg{} opens the possibility of using \alg{} with particular cache replacement algorithms in appropriate situations, for instance when run off of SSDs \cite{Tang2015RIPQ}, \alg{} with \FIFO{} may achieve the best performance.

\subsection{Cache Size and Precision} 
\label{sec: caching}
The \alg{} prefetching layer can accommodates most cache replacement algorithms. To focus the discussion, we will hereby focus only on \LRU{} and \mLRU{}. 

\begin{figure}[!tb]
  \centering
  \includegraphics[width=0.8\linewidth]{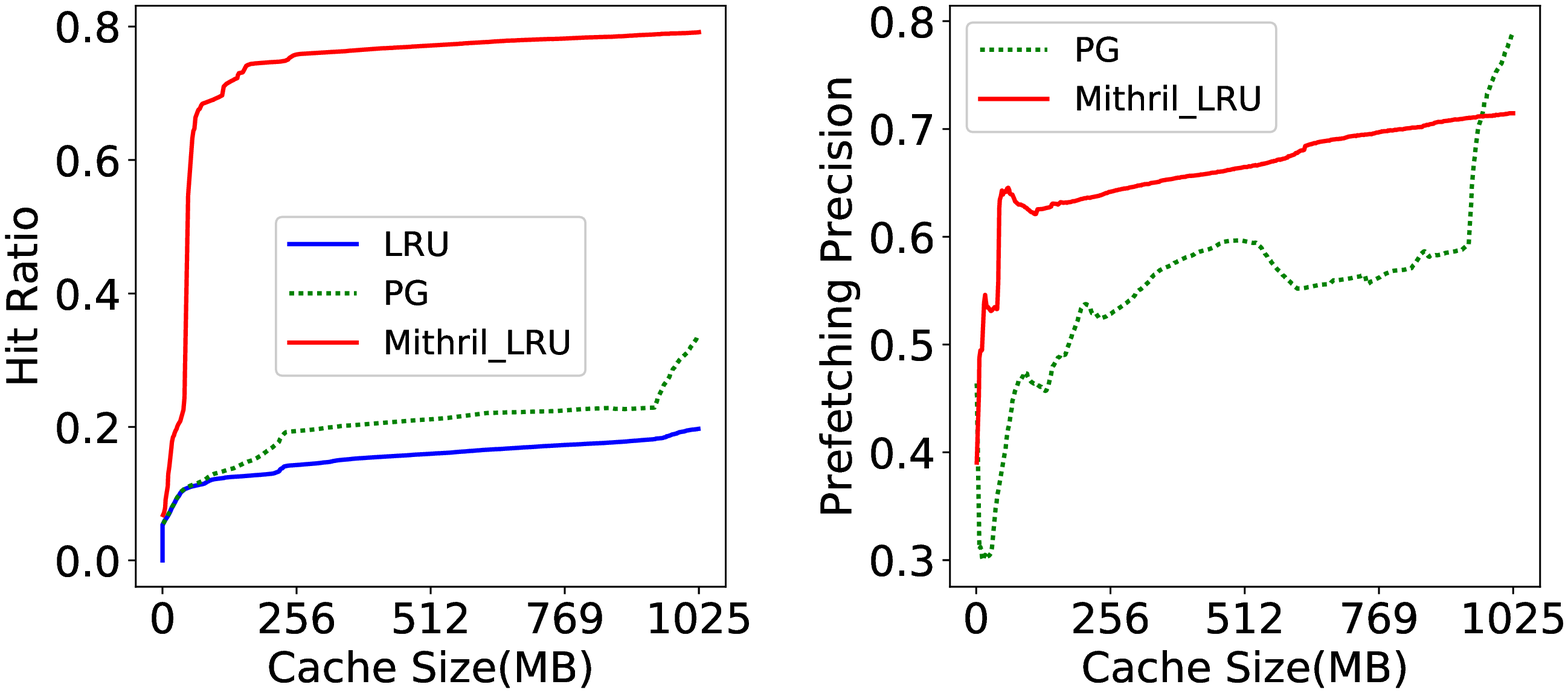} 
\caption{\textbf{Hit ratio curve and prefetching precision of \LRU{}, \PGs{} and \alg{}} \textit{Left: \alg{} outperforms \LRU{} and \PGs{}. Right: The prefetching precision of \alg{} is higher than \PGs{} and both two curves are not monotonic. }}
\label{fig: HRC}
\end{figure}

Our results so far are based on performance at a single cache size. We now show the performance of \alg{} under a range of cache sizes. 
Figure~\ref{fig: HRC} shows the hit ratio curve (HRC) of \LRU{}, \PGs{} and \alg{} along with the prefetching precision of the latter two. Shown in HRC, the performance \PGs{} is always better than \LRU{}, and as the cache size increases, the gap between \PGs{} and \LRU{} increases due to more space allocated for \textsc{PG's} pair-wise probability matrix. However, the improvement of \PGs{} is limited due to its large matrix. In contrast, \alg{} provides a hit ratio boost even at a small cache size. 

The precision curve of \PGs{} has several peaks and troughs because the size of its comprehensive conditional probability matrix depends on cache size. As the cache size increases, the matrix size grows. However, precision may not benefit from the increasing probability matrix size due to wrong new predictions. 
Similarly, the precision curve for \alg{} is also not monotonic, especially with a small cache size, due to the eviction of prefetched blocks before being requested. 
When comparing the prefetching precision of \PGs{} and \alg{}, we see that, in most situations, \alg{} has better precision than \PGs{}. 



\subsection{Effects of Parameters} 
\label{sec: params}

\Alg{} uses several parameters
that now investigate in isolation in terms of impact on hit ratio and prefetching precision using a representative CP trace (w94). 

\begin{figure}[!t]
\begin{subfigure}{0.495\columnwidth}
  \centering
  \includegraphics[width=1\linewidth]{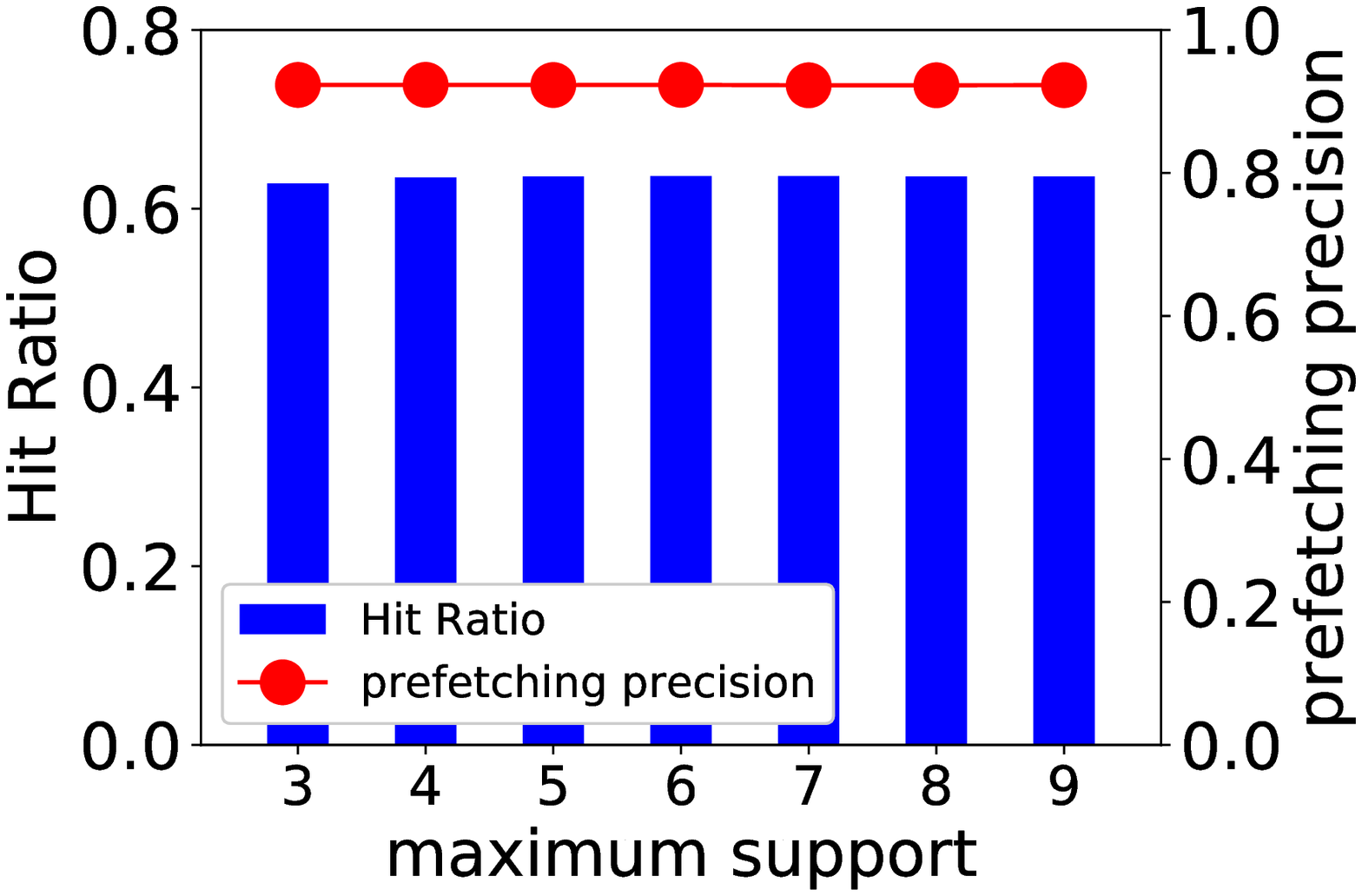} 
  \caption{\maxSup{}}
  \label{fig: paramMaxSup}
\end{subfigure}
\begin{subfigure}{0.495\columnwidth}
  \centering
  \includegraphics[width=1\linewidth]{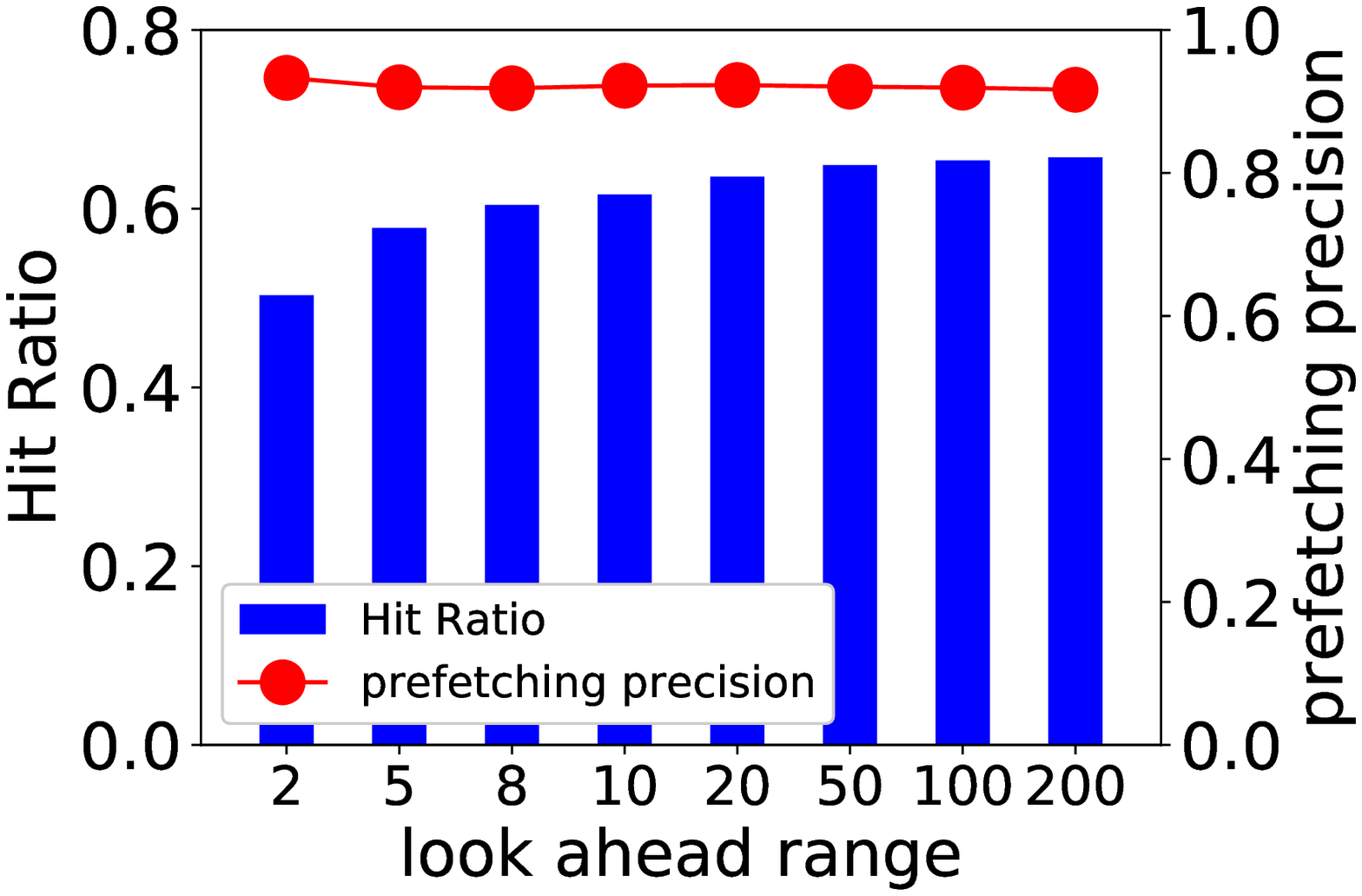} 
  \caption{\lookaheadWindow{}}
  \label{fig: paramLookaheadWindow}
\end{subfigure}
\begin{subfigure}{0.495\columnwidth}
  \centering
  \includegraphics[width=1\linewidth]{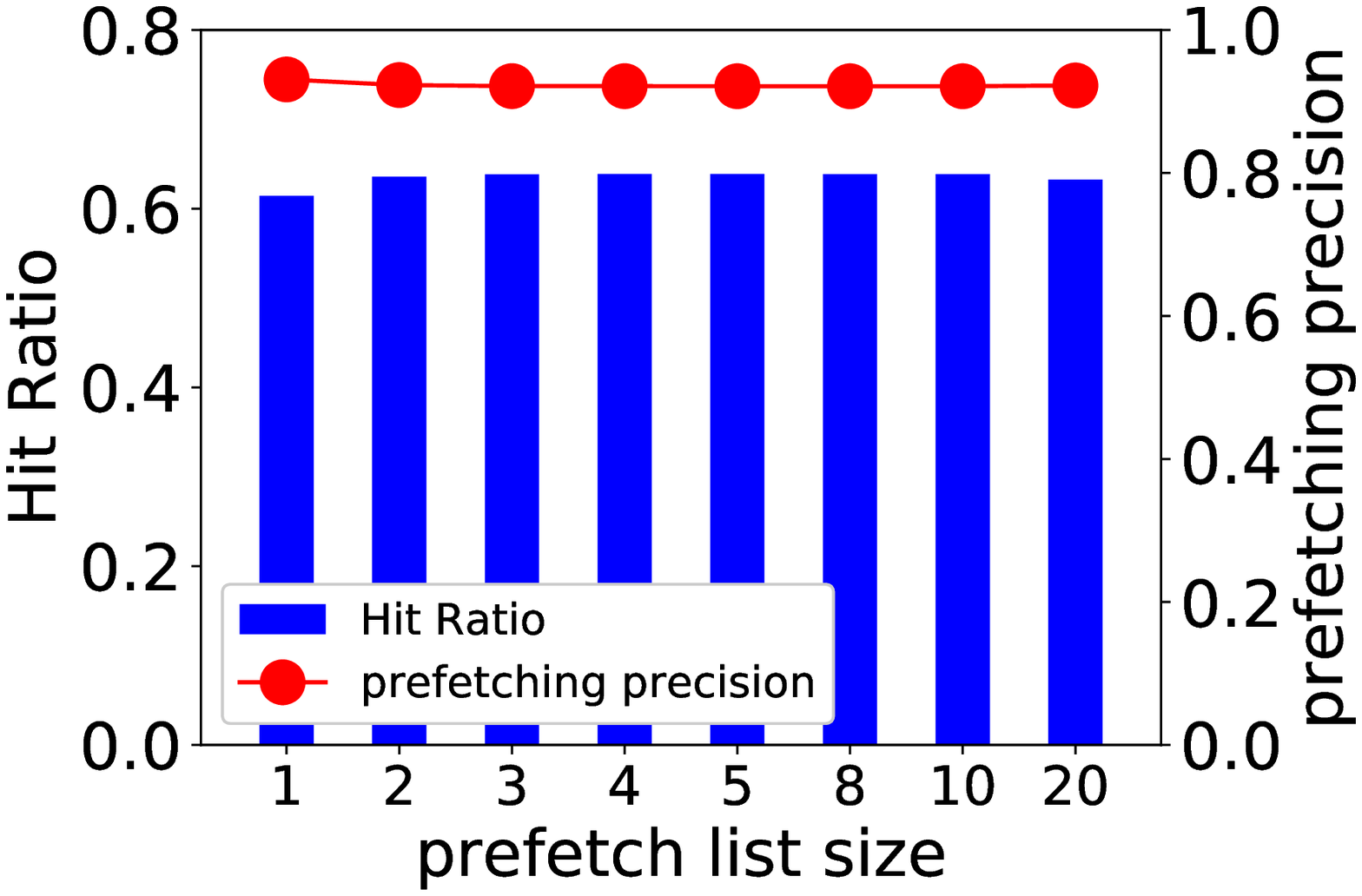} 
  \caption{\prefetchingListSize{}}
  \label{fig: paramPrefetchListSize}
\end{subfigure}
\begin{subfigure}{0.495\columnwidth}
  \centering
  \includegraphics[width=1\linewidth]{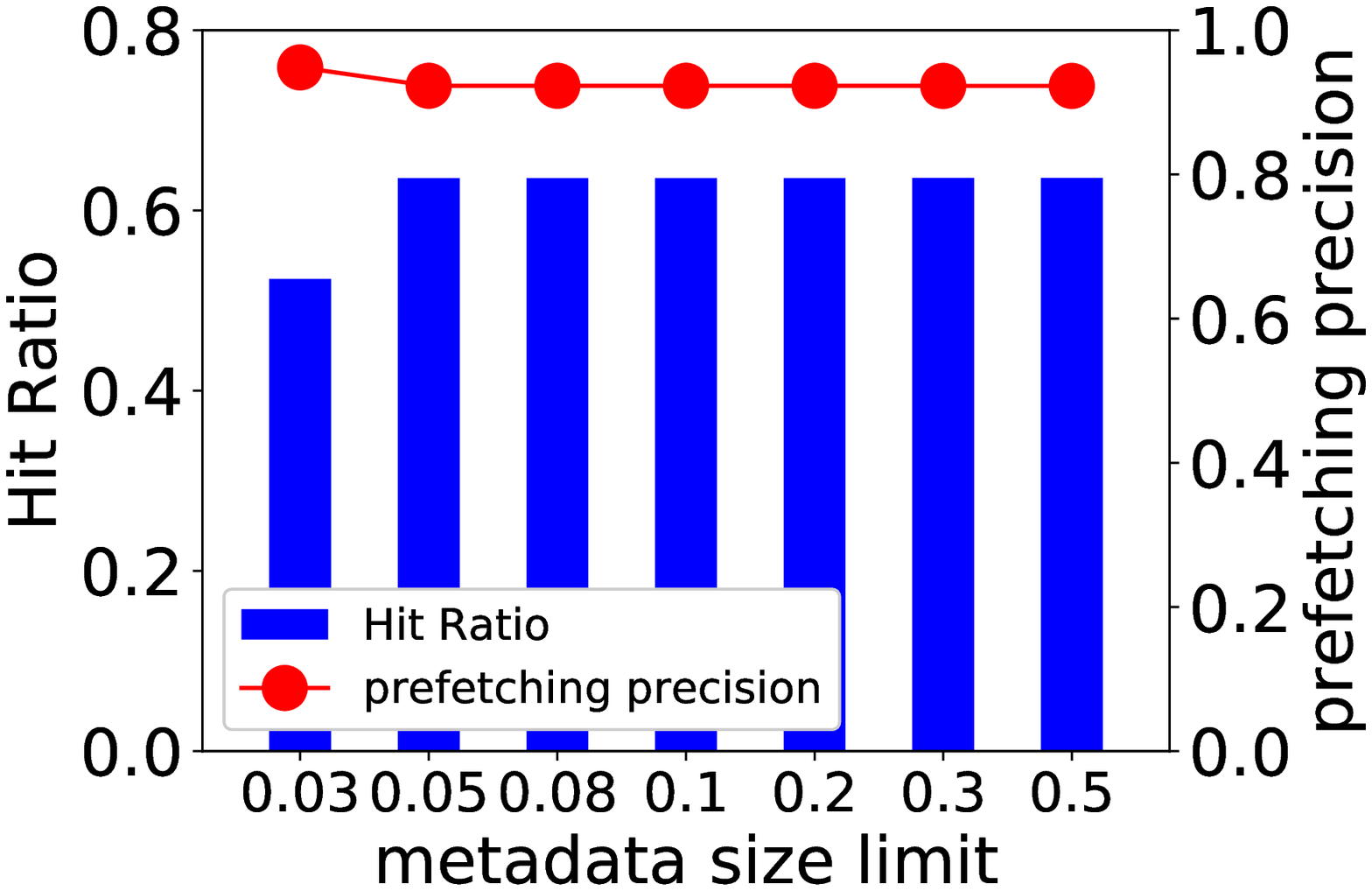} 
  \caption{\maxMetaDataSize{}}
  \label{fig: paramMaxMetaDataSize}
\end{subfigure}
\begin{subfigure}{0.495\columnwidth}
  \centering
  \includegraphics[width=1\linewidth]{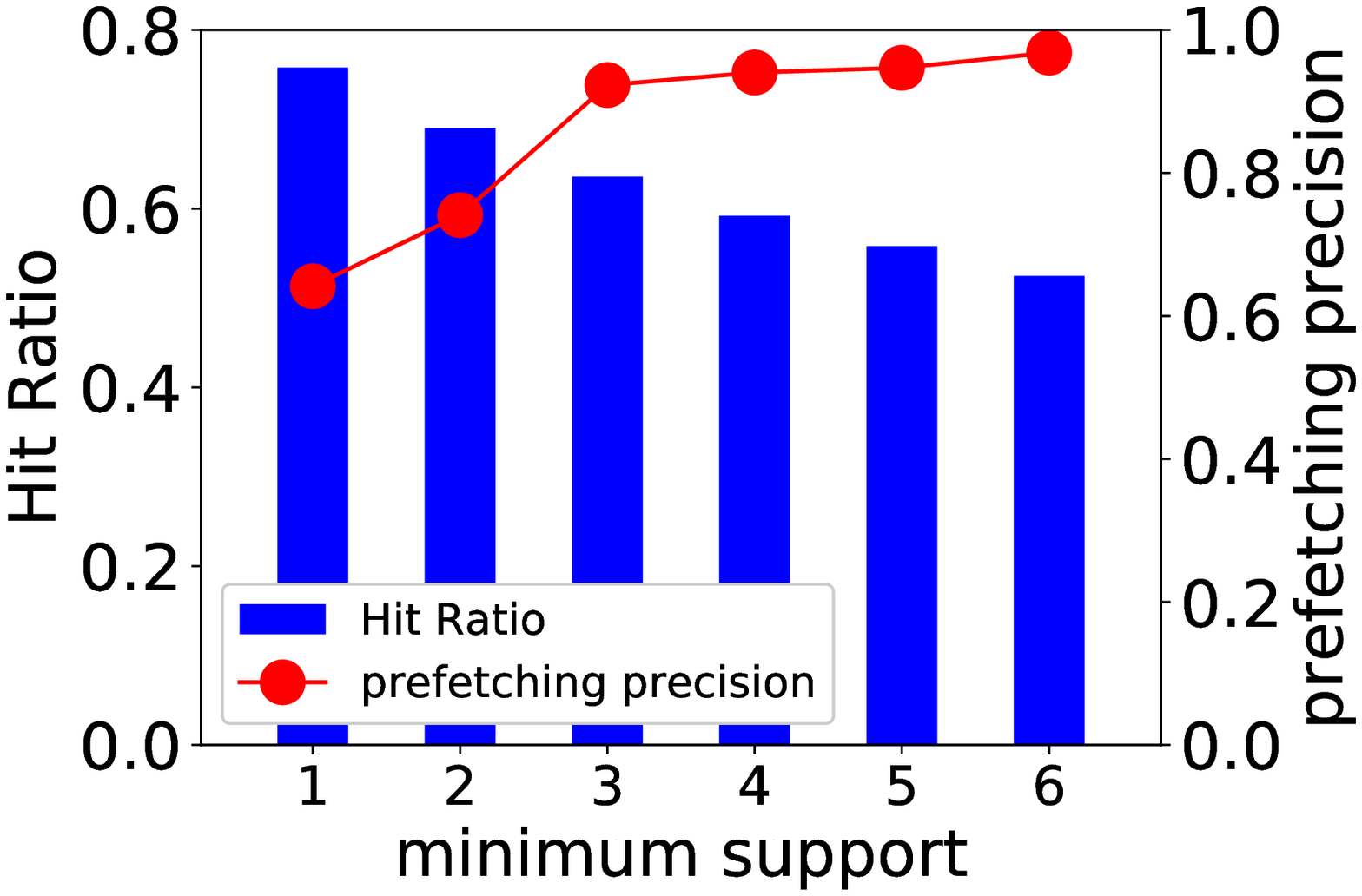} 
  \caption{\minSup{}}
  \label{fig: paramMinSup}
\end{subfigure}
\begin{subfigure}{0.495\columnwidth}
  \centering
  \includegraphics[width=1\linewidth]{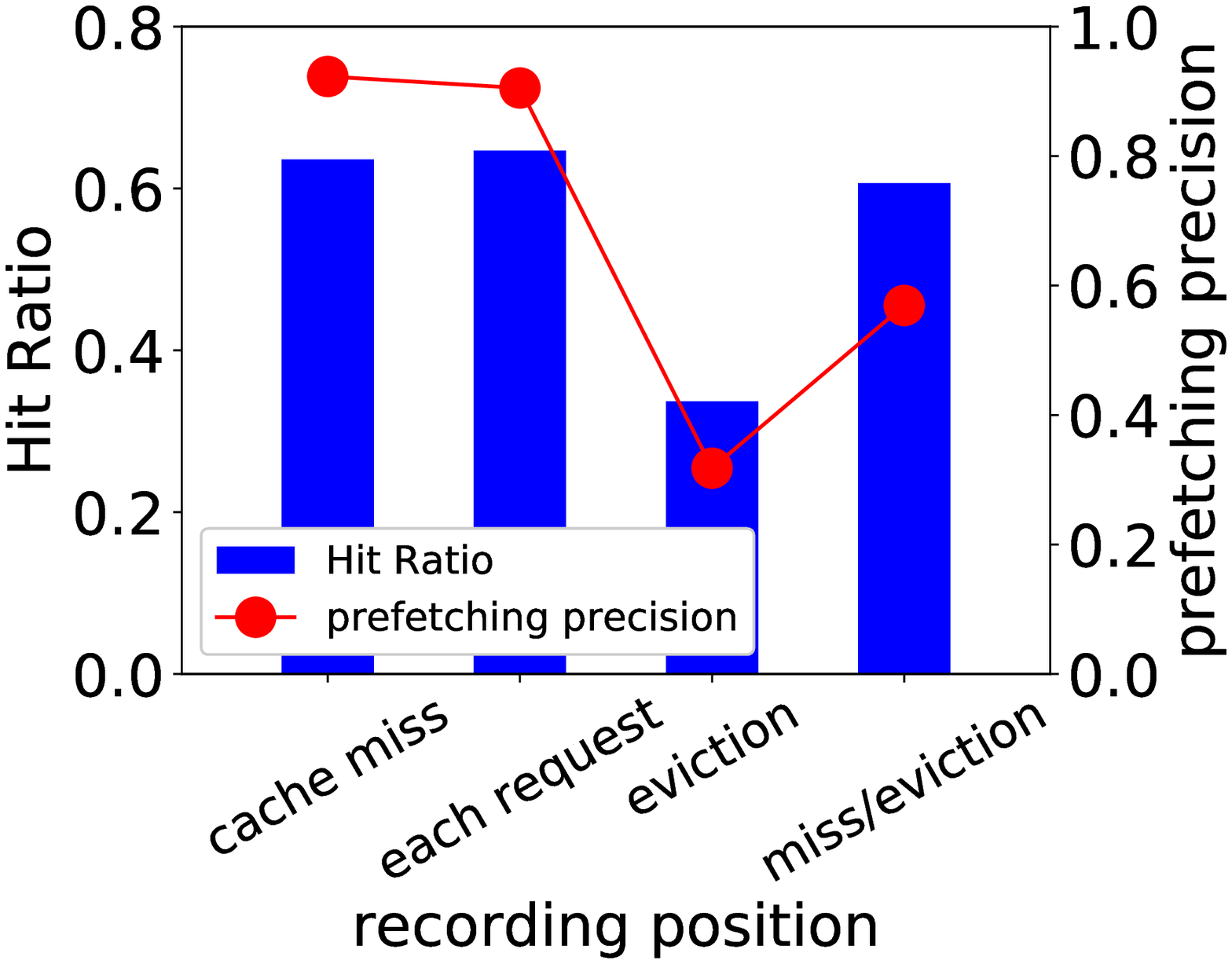} 
  \caption{effect of recording}
  \label{fig: paramRecording}
\end{subfigure}

\caption{\textbf{Effect of parameters in \alg{}.} \textit{}} 
\label{fig: params}
\end{figure}

\textbf{\MaxSup{}} decides the maximum allowed degree of hotness of a block. This is decided by considering the row length of the mining table. If a block is requested more than $S$ times before mining, it gets kicked out as a frequent block. As shown in Figure~\ref{fig: paramMaxSup}, $S$ has a small effect on hit ratio and prefetching precision since most of the frequent blocks are already filtered out by an underlying caching layer. Recall that \alg{} records blocks only during cache misses. 

\textbf{\LookaheadWindow{}} decides the maximum allowed timestamp difference for two blocks to be considered associated. It is obvious that $\Delta$ should be a parameter related to the number of concurrent running processes. If too large, non-associated block pairs will be mistaken as associated, thus increasing the false positive rate. On the other hand, being too small will result in many associations being ignored and thus a high false negative rate. 
As shown in Figure~\ref{fig: paramLookaheadWindow}, when $\Delta$ is small, as $\Delta$ increases, the hit ratio increases substantially, while prefetching precision decreases slightly. After certain threshold, further increasing $\Delta$ will not increase hit ratio. This is because the best $\Delta$ should relate to the number of concurrent running applications, the given trace shown in the figure has its best $\Delta$ around 50. 

\textbf{\PrefetchingListSize{}} determines the space that can be used for storing associated blocks, which is the row length of the prefetching table. Recall that when more than $P$ associated blocks are discovered, the old blocks are replaced in a FIFO manner. Figure~\ref{fig: paramPrefetchListSize} shows that increasing $P$ dramatically reduces prefetching precision because a large $P$ means stale associations are also stored for prefetching. On the other hand, the hit ratio first increases and then decreases with an increasing $P$. We notice that setting $P$ as 2 gives an acceptable trade-off between hit ratio and precision across the various datasets we considered. 

\textbf{\MaxMetaDataSize{}} decides the maximum space \alg{} can use for the recording table, mining table and prefetching table. 
As illustrated in Figure~\ref{fig: paramMaxMetaDataSize}, if $M$ is too small, there are not enough spaces for the prefetching table, dramatically reducing the effect of \alg{}. After a threshold, further increasing $M$ won't increase the hit ratio. However, setting $M$ too large in situations that \alg{} does not have good performance will waste space which should be used for caching. We thus recommend a default value of 10\% of the entire cache space based on traces we have tested. 

\textbf{\MinSup{}} has the largest effect on the performance of \alg{}. It decides when a request is ready for mining and is the row length of the recording table. In Figure~\ref{fig: paramMinSup}, we can see that increasing $R$ will increase prefetching precision, while reducing the hit ratio. Two requests are required to appear closely $R$ times to be considered associated, and when we have a larger $R$, the requirement for being associated is stricter, which diminishes the number of associations and grows the confidence of discovered associations. 

\textbf{Different recording locations} also have a large effect on the performance of \alg{}. As mentioned in Section~\ref{sec: design}, we record only at cache misses, which reduces computation by recording only the most important information. 
As shown in Figure~\ref{fig: paramRecording}, besides recording 
a) at cache miss, we can also record 
b) when a block is evicted from cache, 
c) at cache miss and eviction, or 
d) each time a request arrives. 
Using c) and d) usually give more information to \alg{} at a cost of more computation. In other words, we can trade CPU cycles for potentially better hit ratio and precision. 
As we observe across the traces, recording at evictions (b) usually cannot provide good performance; recording at evictions and misses (c) occasionally provides similar performance to the other two approaches a and d, but most of the time only slightly better than recording at evictions (b). 
In contrast, recording at the arrival of each request (d) usually gives the best performance with the highest precision. As an alternative, recording at cache misses (a) can greatly reduce the overhead of \alg{}, while, as we have evaluated in most traces, it provides less than a 10\% performance loss  compared to recording at each request. 


\subsection{Real System Performance} 
\label{sec: microBenchmark}

\begin{figure}[!tb]
\begin{subfigure}[width=\linewidth]{\linewidth}
  \centering
  \includegraphics[height=1.1\linewidth, angle=-90]{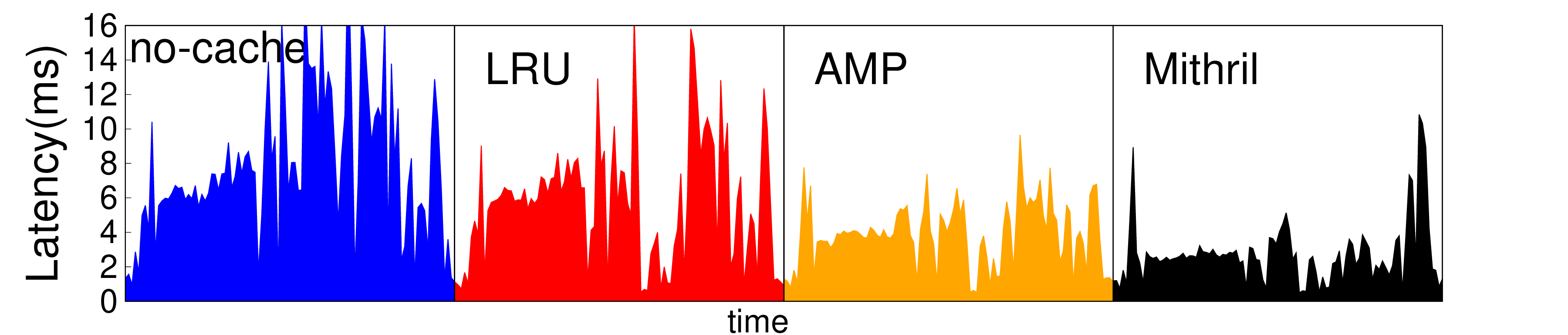} 
\end{subfigure}
\begin{subfigure}[width=\linewidth]{\linewidth}
  \centering
  \includegraphics[height=1.1\linewidth, angle=-90]{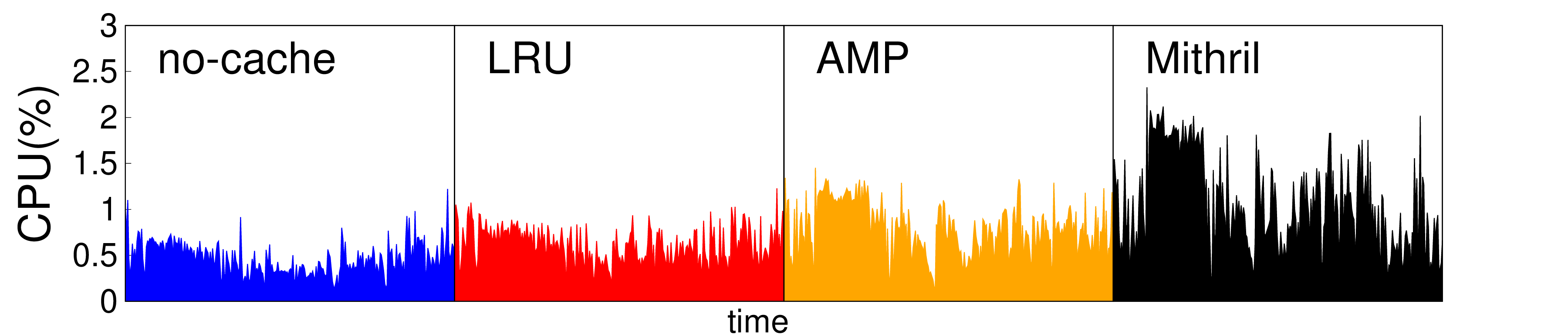} 
\end{subfigure}

\caption{\textbf{Latency and CPU usage of using no cache, \LRU{}, \amp{} and \mLRU{}.} 
\textit{
On the top, each latency point is the average latency of 40000 requests. At the bottom, it shows the relatively increased CPU usage of \alg{} due to mining and prefetching, compared to \LRU{} and \amp{}, the increase is less than 1\%.}}  
\label{fig: LatencyFig}
\end{figure}

\textbf{Latency.} A high hit ratio may not mean low latency in a real system because of factors such as CPU overhead and late prefetch. Especially for a history-based prefetching, the cost of prefetching a random block is large. In Figure~\ref{fig: LatencyFig}, we justify the overhead compared with benefit. It shows the latency of four approaches on CP trace w94: using no cache, using \LRU{} cache, using \amp{} and using \LRU{} cache with a \alg{} prefetching layer. 
\emph{Compared to no cache, \LRU{} reduces average latency by more than 26\%, especially at the peaks, where the no-cache system shows a high latency. Using a sequential prefetcher \amp{}, the latency further decreases by 32\% over \LRU{} on average, whereas  \alg{} with \LRU{} reduced latency by 52\% over \LRU{}. }

\textbf{Late prefetches.} Although latency reduction due to \alg{} prefetching is evident, we also see that 22.4\% of prefetches are late, which means the arrival of prefetched blocks happen after the time they are requested. Late prefetches affect the performance of \alg{} by wasting one disk read unless caught by the disk controller. 

\textbf{\alg{} warm up time.} In Figure~\ref{fig: LatencyFig}, focusing on the first 5\% percent of the requests in a system with \alg{}, we can see the there is no latency reduction at beginning, and latency decrease as time goes from 0\% to 10\%. The decrease occurs because \alg{} needs sufficiently many requests for warm-up before it conducts mining and prefetching. 

\textbf{Existence of latency peaks.} \alg{} does not eliminate all latency peaks. The peaks stem chiefly from two phenomena: they are due to long disk rotational latency or a burst of requests, or a mix of these aspects. When the peaks occur due to long disk rotational latency, \alg{} can effectively reduce latency by prefetching. One extreme case would be if each block request demands the disk to rotate half way to retrieve the content, causing peaks in a system without \alg{}. However, in systems with \alg{}, associations between these requests would be unveiled and harnessed. In other words, \alg{} would prefetch associated block into the cache ahead of its request time, thus lowering latency. On the other hand, if the latency peak is caused by a large number of outstanding I/Os \cite{Gulati2011Pesto}, \alg{} provides less benefit because issuing prefetches only increases the burden on the disk. Consequently, not all latency peaks can be removed by \alg{}. 

\textbf{CPU usage.} \alg{} is based on approximate association mining, which might be CPU-intensive. As shown in the figure, we see some CPU consumption increase for \alg{}, however, the increase is minor and within the limits afforded by many storage systems. 

\subsection{\alg{} Analysis} 
\label{sec: behavior}
\begin{figure}[!ht]
  \centering
\begin{subfigure}{0.38\textwidth}
  \includegraphics[width=1\linewidth]{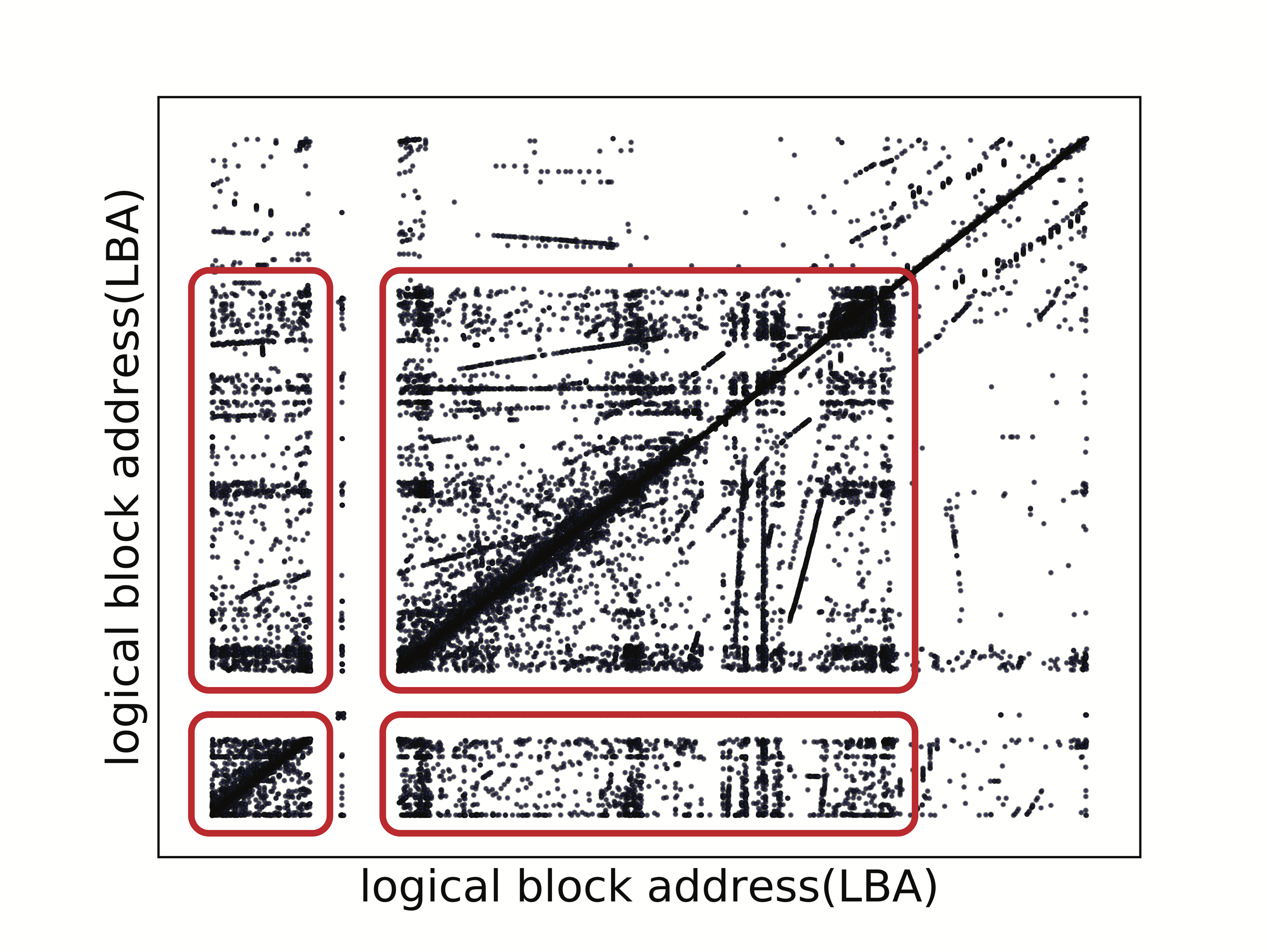} 
  \caption{Associations discovered by \alg{}}
  \label{fig: association}
\end{subfigure}
\begin{subfigure}{0.48\linewidth}
  \centering
  \includegraphics[width=1\linewidth]{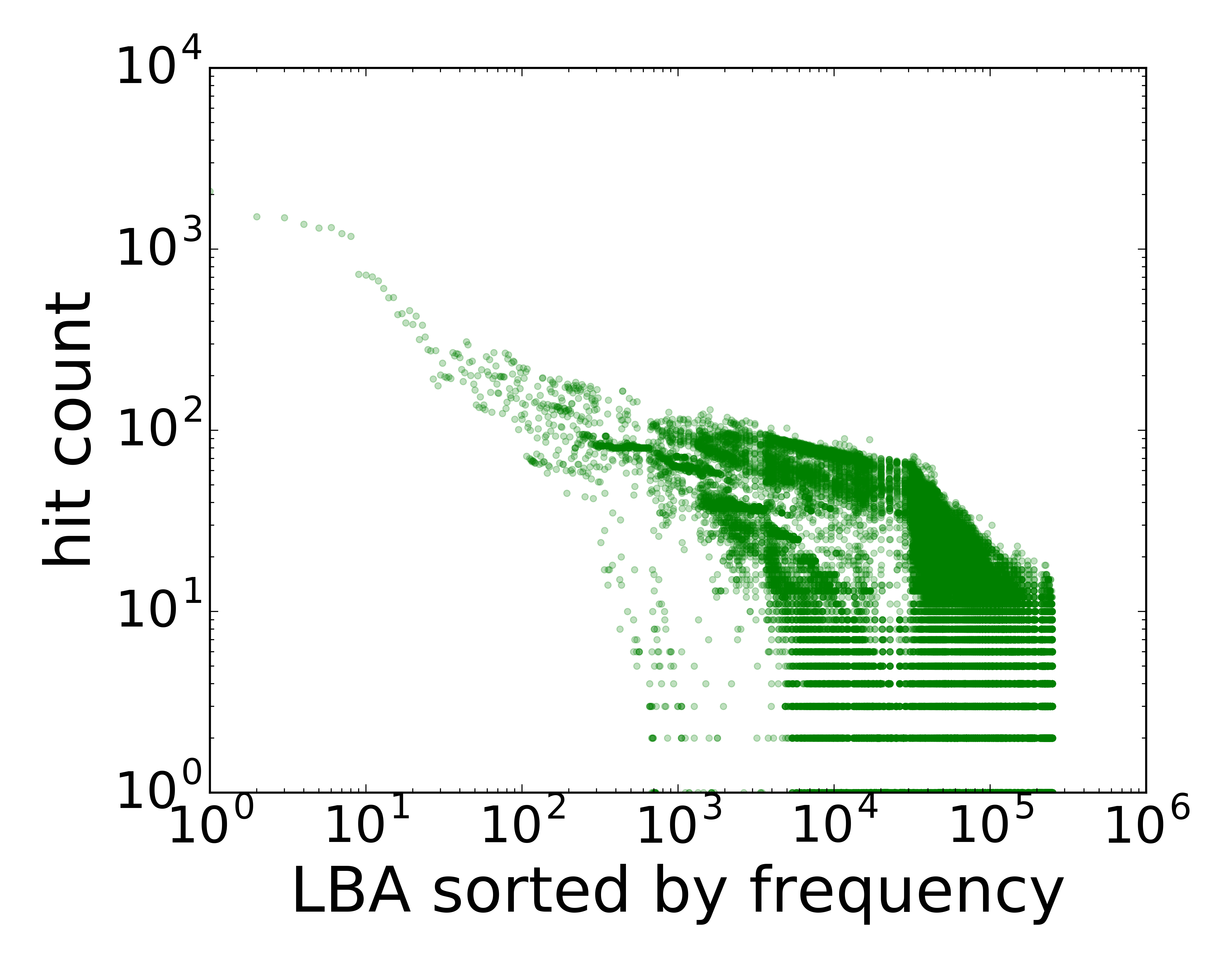} 
  \caption{Hit count in \LRU{}} 
  \label{fig: hitcountLRU}
\end{subfigure}    
\begin{subfigure}{0.48\linewidth}
  \centering
  \includegraphics[width=1\linewidth]{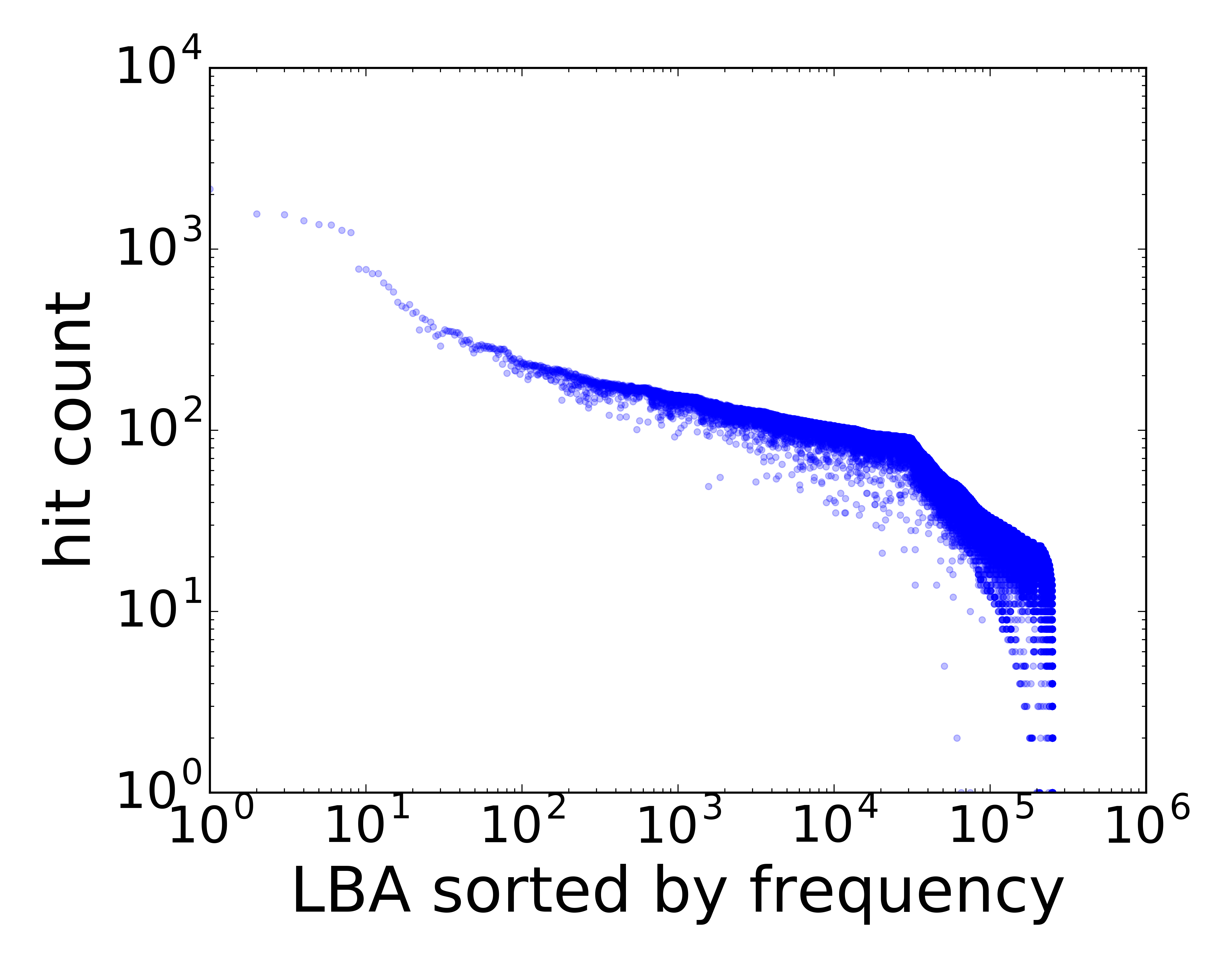} 
  \caption{Hit count in \alg{}} 
  \label{fig: hitcountMithril}
\end{subfigure}    

\caption{\textbf{\alg{} Analysis.} \textit{a): associations discovered by \alg{} contains both sequential associations and non-sequential associations. The four rectangular areas in the figure may represent two major applications that interact with each other. b), c): hit count of blocks sorted by frequency in original trace illustrates \alg{} is able to capture mid-frequency blocks, while \LRU{} cannot. }} 
\label{fig: behavior} 
\end{figure}

In this section, we analyze the behavior of \alg{} underlying its performance. Figure~\ref{fig: association} shows the associations discovered by \alg{} after a full trace run. Both horizontal and vertical axes are logical block addresses (LBA): if two blocks $b_x$ and $b_y$ are associated, a dot is placed at point ($x$, $y$). The association plot clearly shows that \alg{} not only discovers sequential block associations, denoted by the diagonal in the graph, but also many non-sequential block associations. 

As mentioned earlier, \alg{} is designed to catch the mid-frequency blocks since frequent blocks are captured by the underlying caching layer and rare blocks are by definition not worth chasing after. Figure~\ref{fig: hitcountLRU} and Figure~\ref{fig: hitcountMithril} show the hit count obtained by \LRU{} and \alg{}; the horizontal axis is sorted by the frequency of blocks in the original trace. 
\LRU{} gets cache hits on most of the frequent blocks (left part of the figure). For mid or low frequency blocks, \LRU{} shows a bushy image because whether \LRU{} can catch a mid or low frequency block depends on if the block shows small-range locality. If a block shows small-range locality, it can be caught by \LRU{}. For example, if a block is accessed only twice throughout the trace and the two accesses are just separated by a few requests, then it will be captured by \LRU{}. However, if its two accesses are far away from each other, then it won't be captured by \LRU{}. For \alg{}, besides high-frequency blocks being captured, mid-frequency blocks can also be captured because \alg{} can predict its access ahead of time. As shown in the figure, \alg{} has higher hit counts for most blocks in the mid-frequency range. 
These two figures illustrate the crux of why \alg{} provides a high hit ratio: \emph{it discovers sequential associations and non-sequential associations, capturing the mid-frequency blocks that tend to be ignored by common cache replacement policies.}

\section{Conclusion}
\label{sec: concl}
Storage systems increasingly rely on effective caching layers to sustain mounting demands for performance.
We proposed a novel general history-based prefetching layer, \alg{}, to supplement the caching layers.
\alg{} is based purely on the logical timestamp of cache requests without any extra hints, making it easy to use and integrate into existing systems. We evaluated \alg{} on 106 week-long CP traces and 29 70-day-long MSR traces of real storage systems in terms of the hit ratio. 
Our experimental results suggest that \alg{} is lightweight compared to other history-based approaches, and provides 36\% greater hit ratio over the ubiquitous \amp{} sequential prefetching algorithm at modest costs.  

Our work opens a door for combining effective cache replacement algorithms with \alg{} to create a low-overhead caching strategy for capturing often overlooked mid-frequency items and bolster cache performance in today's cloud storage systems.


{\footnotesize \bibliographystyle{acm}
\bibliography{ref}}

\end{document}